%% file: Main.tex
\documentclass[sigconf, screen, 10p, anonymous=False,authorversion,nonacm]{acmart} 
\AtBeginDocument{%
  \providecommand\BibTeX{{%
    \normalfont B\kern-0.5em{\scshape i\kern-0.25em b}\kern-0.8em\TeX}}}

\usepackage{enumitem}
\setlist[itemize]{noitemsep, topsep=0pt}

\usepackage{adjustbox}
\usepackage[linesnumbered,algo2e,ruled]{algorithm2e}
\usepackage{multirow}
\usepackage{booktabs,subcaption,dcolumn}
\usepackage{tikz}
\usepackage{tabularx}
\usepackage{threeparttable}
\usepackage{subcaption}
\usepackage{algorithmic}
\usepackage{graphicx}
\usepackage{textcomp}
\usepackage{xcolor}

\usepackage{balance}

\newcommand\feature[1]{{\footnotesize{\fontfamily{qhv}\selectfont #1}}}
\newcommand\res[2]{\small{$#1$}\tiny{\text{$\pm #2$}}}
\newcommand\bestres[2]{\small{$\mathbf{#1}$}\tiny{\text{$\mathbf{\pm #2}$}}}

\usepackage{array}

\newcommand\MyBox[2]{
  \fbox{\lower0.75cm
    \vbox to 1.7cm{\vfil
      \hbox to 1.7cm{\hfil\parbox{1.4cm}{#1\\#2}\hfil}
      \vfil}%
  }%
}

\ifdefined\showchanges
\newcommand{\changed}[1]{\color{blue}{#1}\color{black}\xspace}
\else
\newcommand{\changed}[1]{#1\xspace}
\fi

\usepackage[strict]{changepage}
\usepackage{framed}
\definecolor{formalshade}{rgb}{1,0.88,0.93}
\definecolor{darkblue}{rgb}{0.84,0.16,0.46}
\newenvironment{formal}{%
  \MakeFramed{\advance\hsize-\width\FrameRestore}%
  \noindent\hspace{-4.55pt}
  \begin{adjustwidth}{}{7pt}%
}
{%
  \end{adjustwidth}\endMakeFramed%
}

\setcopyright{none}
\begin{document}


\title[Insights and Guidelines From Instagram Engagement Mechanisms]{Follow Us and Become Famous! Insights and Guidelines\\From Instagram Engagement Mechanisms}

\author{Pier Paolo Tricomi}
\authornote{Both authors contributed equally to this research.}
\affiliation{%
  \institution{University of Padova, Italy}
}
\email{tricomi@math.unipd.it}
\author{Marco Chilese}
\authornotemark[1]
\email{marco.chilese@trust.tu-darmstadt.de}
\affiliation{%
  \institution{TU Darmstadt, Germany}
}

\author{Mauro Conti}
\affiliation{%
  \institution{University of Padova, Italy}
}
\email{conti@math.unipd.it}

\author{Ahmad-Reza Sadeghi}
\affiliation{%
  \institution{TU Darmstadt, Germany}
}
\email{ahmad.sadeghi@trust.tu-darmstadt.de}

\renewcommand{\shortauthors}{Tricomi and Chilese, et al.}

\input{abstract}

\ccsdesc[500]{Information systems~Social networks}
\ccsdesc[500]{Computing methodologies~Machine learning}
\keywords{Instagram, Engagement, Popularity, Interpretable AI, Social Networks, Post Popularity}

\newcommand\repo{\textcolor{red}{\url{https://anonymous.4open.science/r/FollowUs_Project}}}


\maketitle

\input{Sections/01-Introduction}
\input{Sections/02-Related_works}
\input{Sections/03-Dataset}
\input{Sections/05-Correlations}
\input{Sections/06-Unsupervised}
\input{Sections/07-Guidelines}
\input{Sections/08-Conclusion}


\bibliographystyle{ACM-Reference-Format}


\bibliography{abbr.bib, bibliography}

\end{document}

%% file: abstract.tex
\begin{abstract}
With 1.3 billion users, Instagram (IG) has also become a business tool. IG influencer marketing, expected to generate \$33.25 billion in 2022, encourages companies and influencers to create trending content. Various methods have been proposed for predicting a post's popularity, i.e., how much engagement (e.g., Likes) it will generate. However, these methods are limited: first, they focus on forecasting the likes, ignoring the number of comments, which became crucial in 2021. Secondly, studies often use biased or limited data. Third, researchers focused on Deep Learning models to increase predictive performance, which are difficult to interpret. As a result, end-users can only estimate engagement after a post is created, which is inefficient and expensive. A better approach is to generate a post based on what people and IG like, e.g., by following guidelines.

In this work, we uncover part of the underlying mechanisms driving IG engagement. To achieve this goal, we rely on statistical analysis and interpretable models rather than Deep Learning (black-box) approaches. We conduct extensive experiments using a worldwide dataset of 10 million posts created by 34K global influencers in nine different categories. With our simple yet powerful algorithms, we can predict engagement up to 94\% of F1-Score, making us comparable and even superior to Deep Learning-based method. Furthermore, we propose a novel unsupervised algorithm for finding highly engaging topics on IG. Thanks to our interpretable approaches, we conclude by outlining guidelines for creating successful posts. 

\end{abstract}

%% file: Sections/01-Introduction.tex
\section{Introduction}

People post photos on Instagram (IG) for many purposes, e.g., to convey personal identity, connect with others, or promote worthwhile content~\cite{whyshare}. Getting approval from others is highly rewarding, to the point that engagement metrics (e.g., Likes, Comments, Views) have become addictive, especially for low self-esteem people~\cite{martinez2019likes}. Some people use IG for only a few minutes each day, but for others, e.g., the influencers, it has become a way of life. In short, influencers are people who can influence society. Due to their ability to reach people, companies have used them to market their products~\cite{maslowska2019online}, so much so that influencer marketing is estimated to generate \$33.25 billion in 2022~\cite{instagramuserstatisticstrends_revenue2021, instagramuserstatisticstrends}. Whatever the reason, everyone strives to get as much engagement as possible under their posts, even at the cost of buying it~\cite{zarei2020impersonators,tricomi2022we}. 
For influencers, planning popular posts is a time-consuming and costly activity, that has no guarantees of success. In this regard, a tool that can predict the popularity of a post in advance would be of great interest, especially when sponsored posts are highly remunerated (e.g, Cristiano Ronaldo is paid around \$1 Million for a single post~\cite{conti2022virtual}).

Researchers have proposed algorithms for predicting the popularity of posts, but they are far from perfect (§\ref{subsec:limit}). The first limitation is they measure engagement only in terms of likes, not incorporating stronger forms of interaction or what IG favors, i.e., the number of comments~\cite{instaalg2022}. Then, the lack of a universal dataset for such predictive tasks lead to outcomes based on limited or biased data. Furthermore, these models often make use of Deep Learning (DL) models that may be difficult (or even impossible) to interpret~\cite{rudin2019stop, hee2022explaining}.  
As a result, end-users must \textit{first} create the post going through an expensive and time-consuming process, and \textit{then} assess posts' popularity using such black-box models. 

\begin{figure}[htbp]
    \centering
    \includegraphics[width=\linewidth]{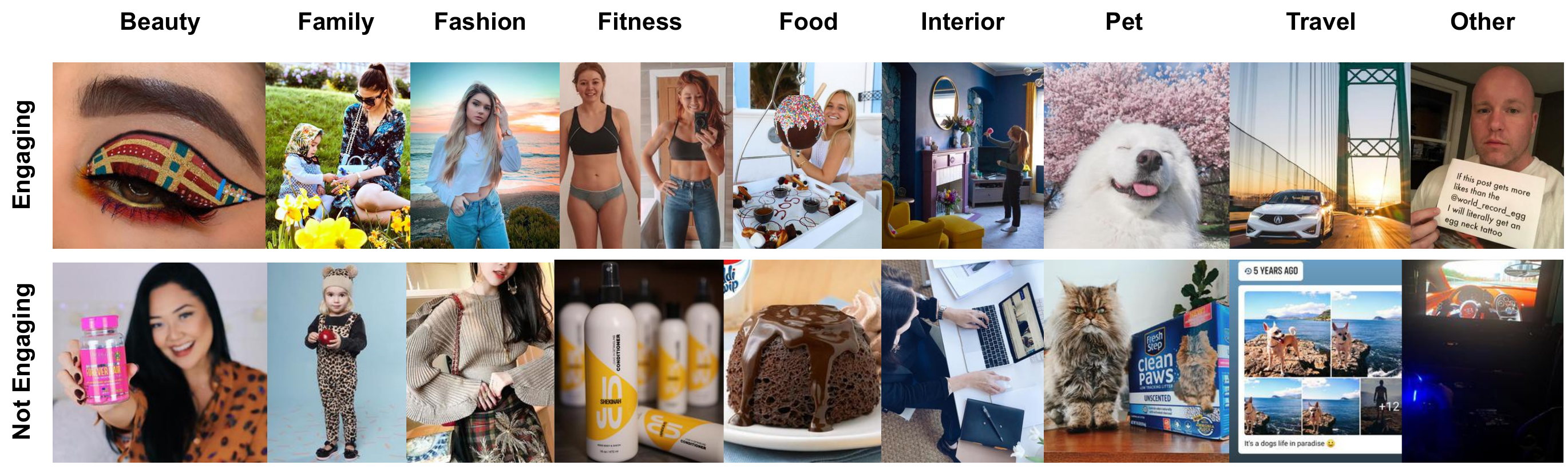}
    \vspace{-2em}
    \caption{Engaging and not engaging posts for each category.}
    \label{fig:examples}
\end{figure}

\textbf{Contribution.}
Our goal is to \textit{understand} and \textit{explain} the underlying mechanisms driving IG engagement. We extract domain-relevant features leveraging the well-known capabilities of DL models, but entrust the prediction to interpretable Machine Learning (ML) algorithms~\cite{molnar2020interpretable}, allowing us to draw guidelines. According to IG recommendation algorithm, we consider both likes and comments as engagement metrics. We conduct extensive experiments on a recent dataset of 10M posts from 34K influencers, demonstrating through statistical analysis that influencer tiers (i.e., their audience wideness) and categories (i.e., the primary topic they cover) are crucial to predict posts' popularity. Figure~\ref{fig:examples} shows engaging and non-engaging posts, which supports the intuition that the characteristics determining engagement differ by category.

Last, we propose a novel unsupervised approach to detect hot topics (i.e., highly engaging) for each category, which overcomes the need for domain knowledge to extract meaningful features. We summarize our contributions as follows:
\begin{itemize}
    \item We analyze the underlying mechanisms of IG engagement, in terms of likes and comments, from a dataset of 10M posts, divided into nine categories and five tiers of influencers, leveraging statistical analysis and interpretable ML algorithms;
    \item We propose an interpretable model to predict posts' engagement and define handy guidelines, exploiting several features extracted by State-of-the-Art (SotA) Deep Learning models;
    \item We propose a novel unsupervised approach for spotting highly engaging topics in each tier and category, considering both visual and textual content;
    \item We release our enriched dataset\footnote{It will be released upon publication.} as a possible baseline for future works.
 \end{itemize}


\textbf{Organization.} §\ref{sec:related} presents related works. §\ref{sec:dataset} describes the dataset and preliminary assessments, while the engagement prediction and interpretation are conducted in §\ref{sec:prediction}. The hot topic detection appears in §\ref{sec:unsupervised}, and the final guidelines are provided in §\ref{sec:guidelines}. §\ref{sec:conclusion} concludes the paper.

\textbf{Transparency.} To promote transparency and reproducibility, we created a repository\footnote{It will be made public upon publication.} containing exhaustive details on our study, the source code, and our dataset.

%% file: Sections/02-Related_works.tex
\section{Related Works}
\label{sec:related}

The popularity of IG posts has been mainly assessed by predicting the number of likes they received, usually divided by the number of followers of the posting user, or after a log-scaled transformation. Mazloom et al.~\cite{mazloom2016multimodal} predicted the popularity of brand-related posts by defining engagement parameters important in marketing and using a Support Vector Regression. The authors extended further their work~\cite{mazloom2018category} for different categories such as activities, landscapes, people, and animals.
De et al.~\cite{de2017predicting} trained a Deep Neural Network (DNN) on posts' metadata (e.g., creation date, users tagged, hashtags) to predict the popularity of future posts of an Indian lifestyle magazine.
Similarly, Zhourian et al. ~\cite{zohourian2018popularity} approached popularity prediction both as a regression and classification task, focusing on posts of Iranian business IG accounts.
Rather than predicting popularity in general, Zhang et al.~\cite{zhang2018become} implemented a dual-attention mechanism  to perform user-specific posts' popularity prediction.
Ding et al.~\cite{ding2019intrinsic} tried to isolate the contribution of the visual content by predicting the intrinsic image popularity through a DNN.
Gayberi et al.~\cite{gayberi2019popularity} extracted concepts and objects features using a pre-trained model on Microsoft COCO Dataset~\cite{COCO} and used several Machine Learning algorithms to predict the likes of a post.
Riis et al.~\cite{riis2020limits} extracted visual semantics such as concepts, scenes, and objects through transfer learning and tried to set an explainable baseline for population-based popularity prediction.
Carta et al.~\cite{carta2020popularity} proposed an approach based on Gradient Boosting and feature engineering of users' and posts' metadata to predict popularity in classification fashion. Last, Purba et. al~\cite{purba2021instagram} attempted to create a global dataset of around 20K posts from 16K users, and leveraged features extracted from hashtags, image analysis, and user history, predicting the number of likes over followers using a Support Vector Regression (SVR).

\subsection{Limitations of Existing Literature}
\label{subsec:limit}
In this section, we briefly describe why the past literature in the area is incomplete, and how our work closes such gaps.

\textbf{Incomplete Popularity Metric.}
Prior works focused \textit{exclusively} on the number of likes to measure post popularity, which is outdated and discrepant with the current IG algorithm. 
The IG algorithm was changed in 2021~\cite{instaalg2022} to show users content based on their interests, not just their social graph. The shift to such \textit{recommendation media} changed how posts became popular. The content \textit{must} be engaged with, mainly through likes and comments, so that Instagram spreads it on many users' feeds, and only then it can become popular.
Consequently, it is crucial to consider the number of comments as an indicator of engagement, given they provide a higher users' expression than likes~\cite{aldous2019view}, and thus are more relevant for the IG recommendation algorithm~\cite{instaalg2022}.
As far as we know, we are the first to include comments in our engagement metrics.

\textbf{Limited or Biased Dataset.} 
Since Meta's APIs\footnote{\url{https://developers.facebook.com/docs/instagram-api/}} are limited, there are no public datasets to use as baselines. Most prior works collected their datasets, focusing on limited portions of the population~\cite{de2017predicting, zohourian2018popularity}. Moreover, except for Mazloom et al.~\cite{mazloom2018category}, they do not consider the different categories and tiers of the creators, although, for example, a pictures of a dog and a top model would become popular for different reasons. The influencer tier, instead, was not previously considered. However, the engagement rate of influencers with millions rather than a few thousand followers reaches different levels~\cite{engagement_what_is_useful}, and normalizing the metrics is not sufficient.
In §\ref{subsec:imp_tiers_categ}, we demonstrate that influencer categories and tiers strongly influence engagement metrics ($p$-value<0.001), and thus need to be treated separately to yield accurate predictions. 

\textbf{Poor Results Explanation.} 
As deep learning algorithms and ensemble machine learning algorithms have improved performance, recent works have largely relied on end-to-end black-box models~\cite{zhang2018become,ding2019intrinsic,riis2020limits} rather than extracting specific features to train simple regressors or classifiers~\cite{mazloom2016multimodal,zohourian2018popularity}. While the model is more accurate, it is difficult (or impossible) to understand what has been learned~\cite{rudin2019stop, hee2022explaining,sharma2022detecting}. As extensively demonstrated in the landmark Nature paper by Rudin~\cite{rudin2019stop}, interpretable models \textit{must} be preferred to (complicated) black-box models when explainability is critical. Often, if the problem has structured data and meaningful features, there is no significant difference in performance between more complex classifiers (i.e., DNNs, ensemble methods) and simpler ones.
Furthermore, in our scenario, using a black-box model for post popularity means that the user must create the post first, which can be extremely costly~\cite{conti2022virtual}. Thus, we use an interpretable model (i.e., a Decision Tree) to provide guidelines that can be followed \textit{before} generating a post that wishes to gain popularity.


%% file: Sections/03-Dataset.tex
\section{Dataset \& Preliminary Assessments}\label{sec:dataset}
In the following section we will describe the dataset (§\ref{subsec:dataset_filter}), the engagement metrics (§\ref{subsec:eng_metr}), the importance of dissecting the data in categories and inner tiers (§\ref{subsec:imp_tiers_categ}), and the features we considered and extracted for the study (§\ref{subsec:feature_extr}).

\subsection{Dataset Description}\label{subsec:dataset_filter}
In our work, we utilize the dataset proposed by Kim et al. \cite{datasetkim2020multimodal} that contains 10,180,500 posts from 33,935 influencers collected in 92 days in total. 
The influencers are divided into nine categories, namely \feature{Beauty}, \feature{Family}, \feature{Fashion}, \feature{Fitness}, \feature{Food}, \feature{Interior}, \feature{Pet}, \feature{Travel}, and \feature{Other}, depending on their content type. Furthermore, we categorize each influencer in the five well-known tiers based on their number of followers~\cite{tricomi2022we}: Nano [1K, 10K), Micro [10K, 50K), Mid [50K, 500K), Macro [500K,1M), Mega [1M, $+\infty$].

Each post is composed of the image, caption, metadata (e.g., publish time, location), and engagement metrics (i.e., the number of likes and comments)\footnote{We did not further process these metrics, e.g., by removing spam comments, since IG algorithm accounts for quantity, and not quality~\cite{instaalg2022}.}.
Similar to previous works~\cite{zohourian2018popularity, riis2020limits,purba2021instagram}, we normalize our target features (likes and comments) dividing them by the number of followers of the post's creator, allowing a fair comparison between posts of different users\footnote{As a convenience, we refer to the normalized numbers simply as likes and comments.}. 
Given that creators' followers were taken only at the end of the collection, we remove posts older than thirty days, a period within the followers' growth remains mostly stable~\cite{instagram_growth}. Moreover, since an IG post engagement growth usually last one to three days~\cite{engagement_what_is_useful}, we exclude posts younger than five days. In the end, our dataset counts 650,118 posts created by 33,935 influencers. Table~\ref{tab:posts-infl} shows the number of posts (and influencers) per tier and category. The small presence of some categories (e.g., food, interior, pet) for very popular influencers is aligned with the actual IG categories distribution~\cite{IGcatdistr}. 

\begin{table}[!ht]
\tiny
\caption{N. posts (influencers) for categories and tiers.}
\label{tab:posts-infl}
\resizebox{\columnwidth}{!}{%
\begin{tabular}{lccccc}
\toprule
\textbf{} & \textbf{Nano} & \textbf{Micro} & \textbf{Mid} & \textbf{Macro} & \textbf{Mega} \\
\midrule
\textbf{Beauty} & 8449 (546) & 7879 (537) & 6998 (387) & 745 (35) & 835 (37) \\
\textbf{Family} & 29744 (1887) & 23432 (1330) & 12740 (674) & 1267 (77) & 2145 (102) \\
\textbf{Fashion} & 49622 (3154) & 82895 (4841) & 68737 (3238) & 8833 (325) & 8987 (355) \\
\textbf{Fitness} & 5060 (301) & 6194 (424) & 6256 (342) & 352 (27)  & 748 (39) \\
\textbf{Food} & 27697 (1511) & 28191 (1440) & 14805 (583) & 936 (25) & 305 (6) \\
\textbf{Interior} & 6461 (373) & 9606 (541) & 5525 (261) & 413 (13) & 404 (7) \\
\textbf{Pet} & 3416 (164) & 4073 (260) & 2929 (153) & 87 (6) & 115 (4) \\
\textbf{Travel} & 24445 (1774) & 19630 (1522) & 13098 (838) & 816 (49) & 540 (27) \\
\textbf{Other} & 73213 (2976) & 38967 (1454) & 31874 (1004) & 4255 (120) & 6399 (166)\\\bottomrule
\end{tabular}%
}
\end{table}



\begin{figure*}%
    \centering
    \subfloat[\centering Likes.]
    {{\includegraphics[width=0.46\linewidth]{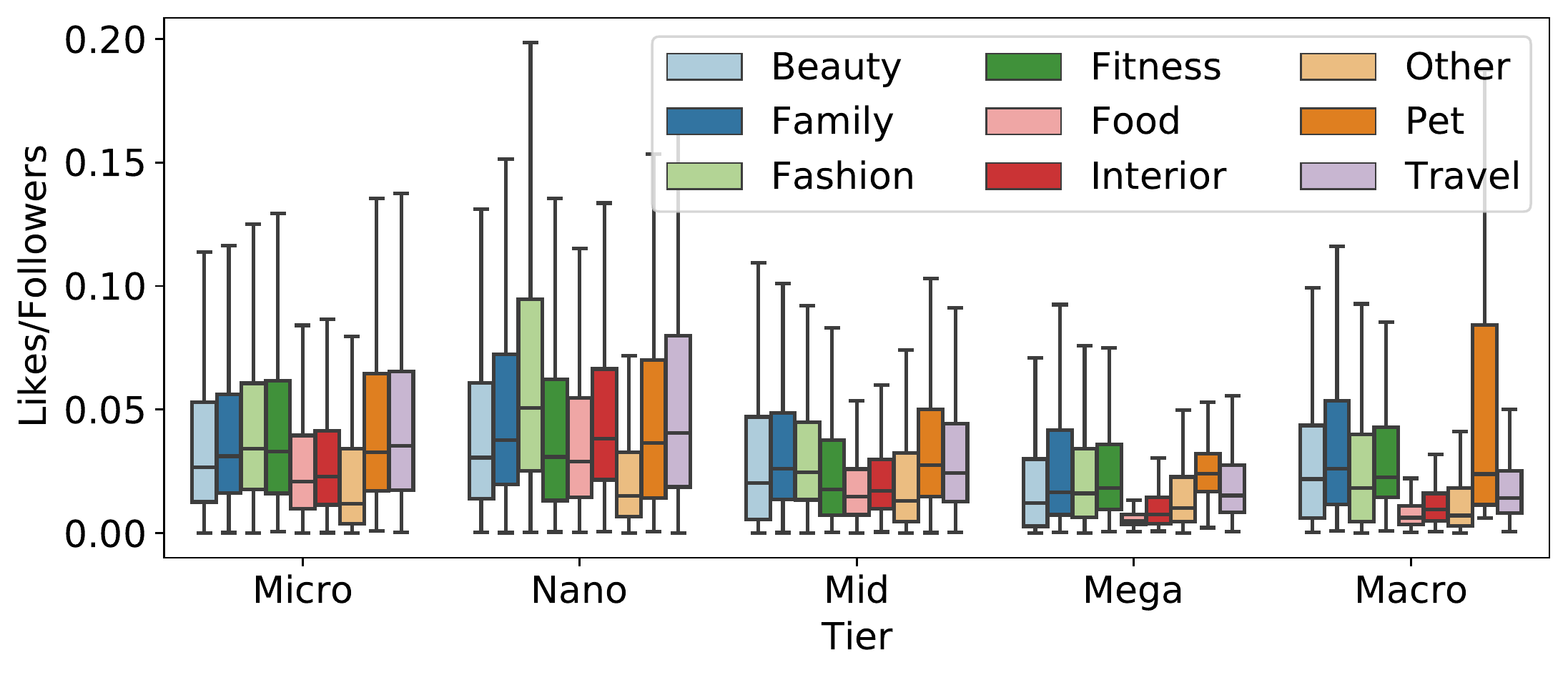} }}%
    \qquad
    \subfloat[\centering Comments.]
    {{\includegraphics[width=0.46\linewidth]{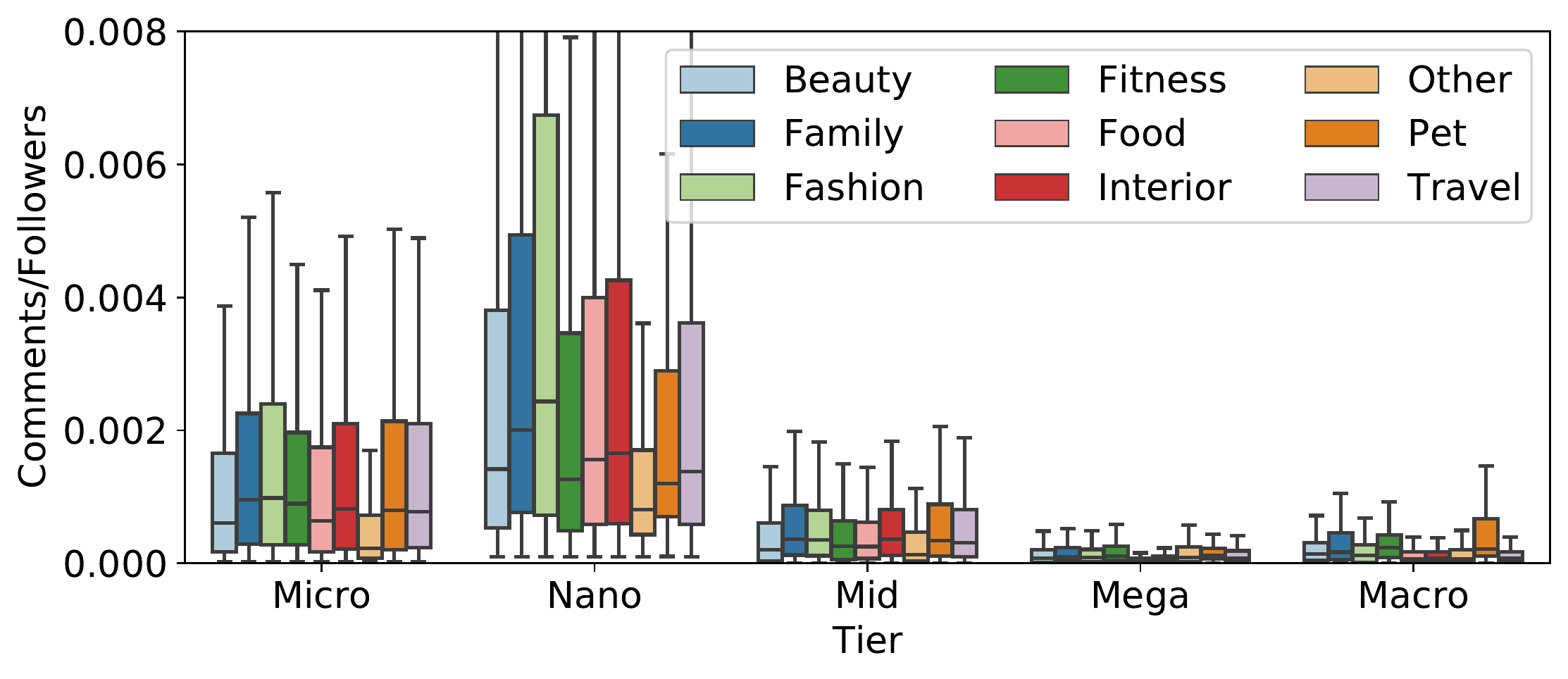} }}%
    \caption{Box plots of Likes and Comments for the different categories and tiers. Note that the y-axes have two different scales, giving the lower number of comments in general.}%
    \label{fig:boxplot}%
\end{figure*}

\subsection{Engagement Metrics: Likes \& Comments}
\label{subsec:eng_metr}
Prior works (§\ref{sec:related}) focused \textit{exclusively} on the number of likes as a popularity metrics. Nonetheless, since 2021, comments have become a crucial engagement metric to make a post popular~\cite{instaalg2022}. Figure~\ref{fig:boxplot} shows the box plots of the likes and comments for every category and tier. 
There are some common trends, but commenting is less frequent than liking. Such discrepancy is justified by the two different levels of public expression they carry~\cite{aldous2019view}. Comments are costly and expose users' opinions more, while likes are almost immediate and instinctive. Hence, a highly-liked post may not receive many comments. 
\changed{Further demonstrating the independence of the two metrics, we calculated Spearman correlation coefficients ($rho$) between the distributions of likes and comments. }
The result ($rho=0.58$, $p$-value < 0.001) shows a moderate correlation between likes and comments, demonstrating that they need to be analyzed separately as two not-so-dependent phenomena. Thus, we consider as engagement metrics $\frac{\#Likes}{\#Followers}$  and $\frac{\#Comments}{\#Followers}$.

\subsection{The Importance of Tiers and Categories}\label{subsec:imp_tiers_categ}

Do the Northern Lights create more engagement than a cute puppy? How about a pineapple pizza in Naples?
As these concepts are incomparable, answering these questions a priori is difficult. Similarly, would people react the same way if a celebrity and a normal person divorced? Most likely not.
Those are just a few examples behind our hypothesis: \textit{influencers' tiers and categories have a significant effect on the engagement metrics.}
To demonstrate this hypothesis, we conduct a Multivariate ANOVA (MANOVA)~\cite{MANOVA}, with category and tier as independent variables, and the likes and comments as dependent ones. By such statistical test, we can determine whether the mean scores of engagements differ between our nine categories and five tiers. Before conducting MANOVA, we normalize the likes and comment distributions as explained in §\ref{subsec:eng_metr}. 
Among the MANOVA results, we adopted Pillai's trace test, which is robust when MANOVA assumptions are violated~\cite{tabachnick2007using}.
Pillai's trace test returned 0.0942 and 0.2646 for category and tier respectively, both with $p$-value < 0.0001. Since the $p$-value is less than the significance level $\alpha = .0001$, we reject the null hypothesis of the MANOVA and conclude that the explanatory variables (tier and category) have a significant effect on the values of the response variables (likes and comments). In particular, the tier resulted contributing more than the category.

\subsection{Features Extraction}\label{subsec:feature_extr}
Starting from the filtered posts of §\ref{subsec:dataset_filter}, we augmented our dataset with  features from each kind of data sources, such as metadata, images, and text, which we now briefly describe.\footnote{The complete list of features is available in our repository.} In the process, we also employed nine SotA DL algorithms.  

\subsubsection{Metadata Features}
The posts' metadata provides information on their ``discoverability''. This term refers to features that increase post visibility, like hashtags and mentions. Hashtags are used to label the post's content, while mentions allow tagging someone in a post, so their followers can reach the source profile. Therefore, we created two counters to keep track of the number of hashtags and tagged users. In addition, we specify whether the post is a video, sponsored, has a location, and time-related information, for a total of 10 features.

\subsubsection{Images Features}
We extract features from images on multiple levels to fully describe the image content, including the scene features, people features, and aesthetic features.

\textbf{Scene features.}
To describe the environment where the picture is set we leverage the Places365 DL model~\cite{places365}. The model can identify up to 365 different places mapped to 3 macro categories (indoor, outdoor natural, outdoor man-made) and 16 micro categories (e.g., shopping/dining, transportation). Moreover, we perform object detection of 80 different classes mapped in 12 categories using Faster R-CNN MobileNetV3~\cite{NEURIPS20199015}) trained on MS COCO dataset~\cite{COCO}, counting the objects belonging to each category.

\textbf{People Features.}
Using RetinaFace \cite{retinafacepaper} we perform faces boundaries detection and then estimate the age and gender~\cite{agegender} of the detected people. For each post, we save the number of females and males, and min, max, mean and standard deviation of people's age. Furthermore, guided by the well-known impact of nudity in advertising~\cite{sherman2006influence}, we perform nudity detection using NudeNet~\cite{nudenet} for \feature{Beauty} and \feature{Fashion} categories, in which the main subject is the human body. The model determines whether 16 parts of the body (e.g., breast, belly, feet, buttocks) are exposed.

\textbf{Aesthetic features.}
Taking inspiration from Guntuku et al.~\cite{twittercolordepression}, we derive aesthetic features of the image. In particular, we first extract the percentage of red, green, and blue channels. Then, from the HSV (Hue, Saturation, Value) representation, we obtain the percentage of luminance, warm and cold colors, pleasure, arousal, and dominance scores~\cite{mehrabian1974approach, valdez1994effects}. Furthermore, we leverage Kong et al.~\cite{kong2016photo} model to obtain eleven high-level aesthetic features (e.g., color harmony, motion blur, symmetry of the content).
Last, we extracted the sentiment score conveyed by the image through the model proposed by Campos et al.~\cite{campos2017pixels}.

\textbf{Other features.}
For the \feature{Pet} category, we calculated pets' cuteness scores through a Cute Animal Detector~\cite{cutedetector}. In total, we obtained 80 visual features.

\subsubsection{Text Features}
From the posts' captions we extracted features such as the caption length, the number of Emojis, and their relative sentiment~\cite{10.1371/journal.pone.0144296}. Moreover, we retrieve the sentiment of the whole text leveraging Google Cognitive Services (GCP)~\cite{gcp}, expressed as a score ($\text{Sentiment}_\text{score} \in [-1, 1]$, where $-1$ is negative, $0$ is neutral and $+1$ is positive) and magnitude ($\text{Sentiment}_\text{magnitude} \in [0, +\infty)$), that is representing the strength of the sentiment. We translated non-English text using 
GCP, and obtained five textual features in total.

%% file: Sections/05-Correlations.tex
\section{Predict \& Interpret the Engagement}
\label{sec:prediction}

Through correlation analysis, we uncover features that correlate with engagement. Then, we use interpretable models to predict engagement and develop guidelines for producing engaging content.

\subsection{Correlation Analysis}\label{subsec:correlation}
To determine which features contribute the most to raising engagement, we correlate the features with our two engagement metrics (Likes and Comments). To this aim, we use Spearman’s rank correlation coefficient $r_{s}$~\cite{spearman}. This method offers the advantages of producing feature ranks, being insensitive to outliers, and not requiring any specific normalization of the data. The Spearman's correlation coefficient is based on Pearson's correlation coefficient~\cite{statisticbook} and it is defined as follows. For $n$ observations, the $n$ scores $X_{i}$, $Y_{i}X_{i}$, $Y_{i}$ are converted to ranks as $R({X_{i}})$ and $R({Y_{i}})$, and $r_{s}$ is computed as:

\begin{equation}
r_{s}=\rho_{R(X),R(Y)}=\frac {\text{cov}(R(X),R(Y))}{\sigma _{R(X)}\sigma_{R(Y)}}
\end{equation}
\noindent
where $\rho$ denotes Pearson correlation coefficient but applied to the rank variables, $\text{cov}(R(X),R(Y))$ is the covariance of the rank variables, $\sigma_{R(X)}$ and $\sigma_{R(Y)}$ are the standard deviations of the rank variables. As for Pearson's correlation coefficient, the Spearman correlation values are expressed in the range $r_s \in [-1, 1]$ along with their $\rho\text{-value}$ that express their significance that is higher as much as the value is small.

\begin{figure}[h!]
    \vspace*{-.35cm}
    \hspace*{-5cm}
    \begin{subfigure}{\textwidth}
        \centering
        \includegraphics[scale=.45]{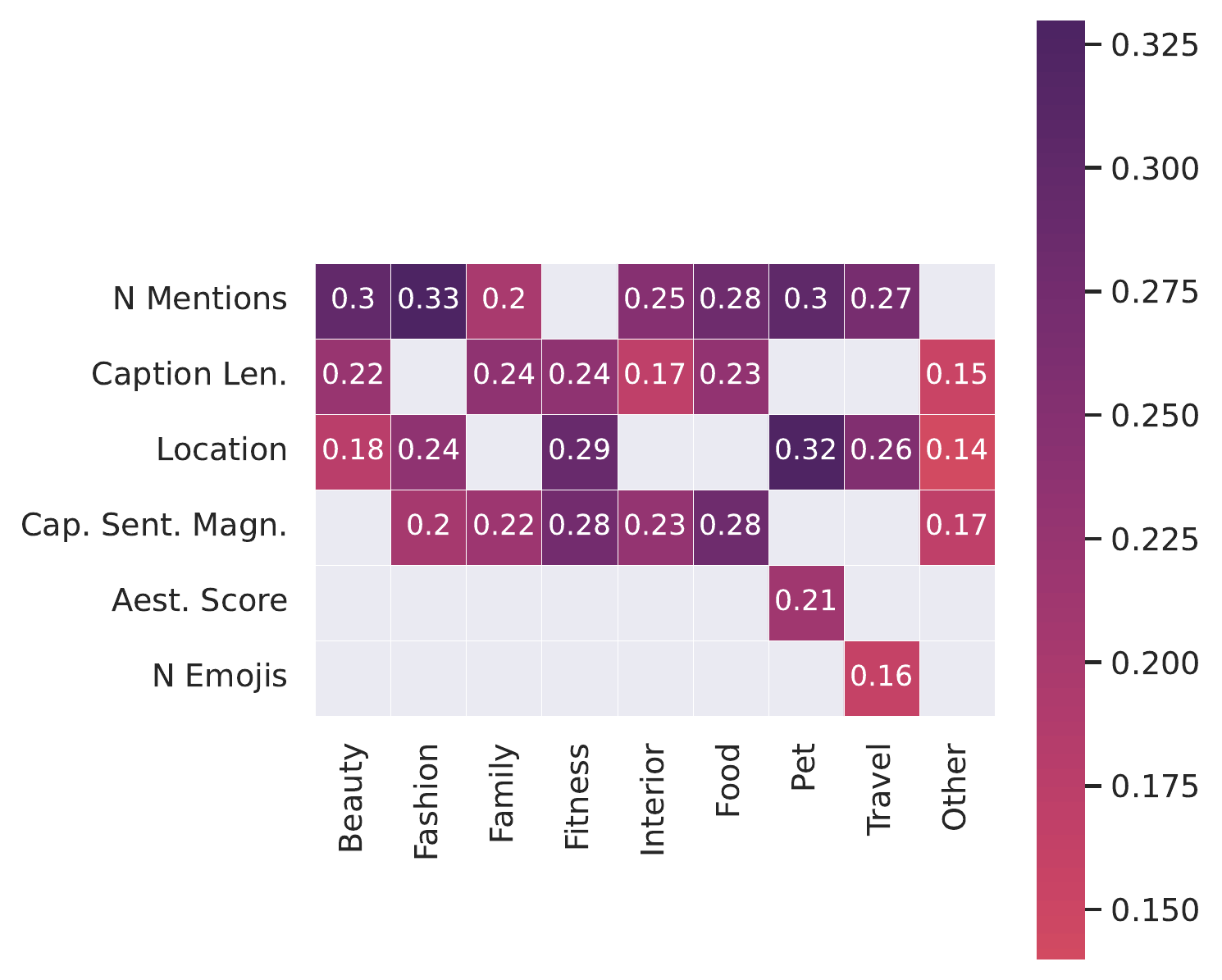}
        \caption{Nano.}
        \label{fig:heatmap_corr_nano_likesFoll}
    \end{subfigure}
    \hspace*{-4.7cm}
    \begin{subfigure}{\textwidth}
        \centering
        \includegraphics[scale=.35]{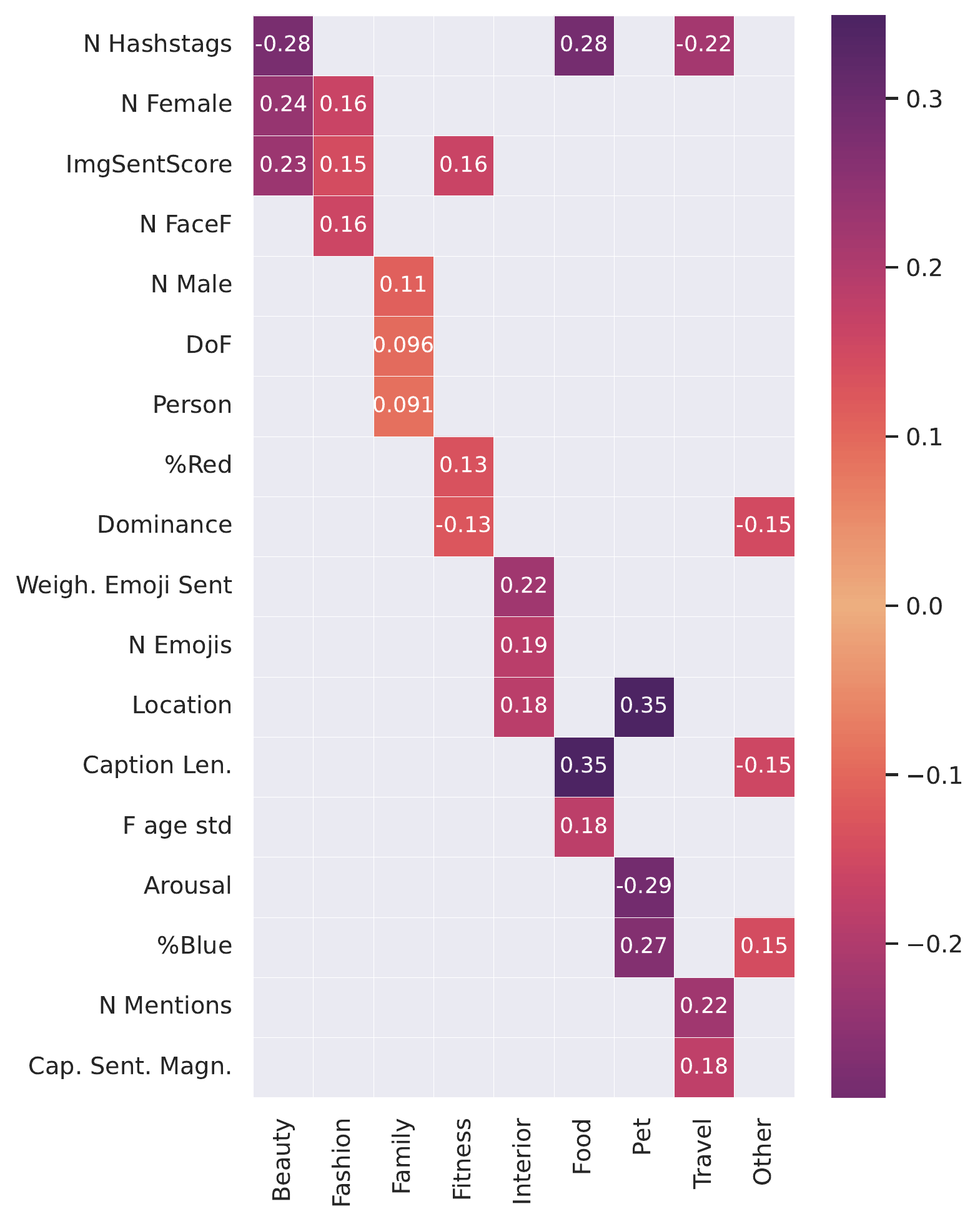}
        \caption{Mega.}
        \label{fig:heatmap_corr_mega_likesFoll}
    \end{subfigure}
    \caption{Top 3 features per absolute correlation value. Comments engagement.}
    \label{fig:correlation_mega_nano}
\end{figure}

For each one of the influencers categories (i.e., \feature{beauty}, \feature{fashion}, etc.) and for each one of the tiers (i.e., nano, micro, etc.) we perform the correlation analysis of the features against the engagement metrics\footnote{We made all the results available in our repository.}.
As the first result, we notice immediately, as depicted in Figure~\ref{fig:correlation_mega_nano}, how the most relevant features in different categories are very similar when the tiers are small (e.g., Nano and Micro). In contrast, behavior becomes category-specific as tier size increases (e.g., Macro and Mega), and this behavior holds for both likes and comments. As we can see, small influencers, or users aspiring to become influencers, use similar strategies in every category. These include the use of many mentions, a long caption, and location tags.

\textbf{Likes Engagement.}
By examining how features correlate to likes, it is possible to observe how the engagement mechanism differs for each type of influencer. Generally, we can notice that the strongest features are related to the images and their content rather than to the text (i.e., the caption), while almost the opposite occurs for the comments. The number of mentions generally has a positive impact on likes, even if their relevance decreases as tiers increase. A similar pattern can be observed in the number of hashtags having a negative effect, which tends to intensify in larger tiers. Availability of the location is very relevant up to the micro tier, after which it becomes category-specific. 

\textbf{Comments Engagement.}
Similarly to the likes engagement, we observe an overall positive correlation for the number of mentions, even though the relevance goes decreasing as the tier increases. Even in this case, the number of hashtags plays an antagonist role in comments engagement, instead the presence of the location field in a post is generally helpful.
This type of engagement benefits from text-specific features such as caption length, sentiment magnitude, and Emoji usage. 

\begin{formal}\textit{
\textbf{Takeaway:} Likes engagement differs from Comments engagement in that they are oriented toward images and captions, respectively. Additionally, low-tier influencers tend to adopt the same strategy to grow, while high-tier influencers exhibit more category-specific characteristics.
}
\end{formal}

\subsection{Engagement Prediction \& Guidelines Methodology}\label{subsec:shaping_guidelines}

\input{Sections/result_table.tex}

Besides explaining which characteristics of Instagram posts build engagement, we also aim to form guidelines for producing the ideal engaging post. Having such guidelines for influencers results in saving time and money consistently since the process for producing a high-engagement post is well-defined. To this aim, we leverage interpretable models, even if this could reduce the overall accuracy. Deep learning models are well known for their capability of solving complex tasks, but by definition, they work as a black box, that we cannot really explain~\cite{rudin2019stop}. For this reason, we decided to utilize Decision Trees (DT)~\cite{quinlan1986induction}. \changed{By training a DT classifier to predict low or high engagement (bottom 0.75 and top 0.25 quantile) we can explain how to produce top engagement posts simply by following the binary classification tree. The paths to reach the top 0.25 quantile leaves represents guidelines for creating high-engagement posts. 
}

\subsubsection{Implementation}
Since influencers behave differently according to their category and tier, as they want to reach a different public, our analysis is performed per each category and tier. With the aim of building an accurate estimator, for each dataset, we fine-tune the Decision Tree Classifier through a Grid Search (cv=5) that is evaluating more than 20K combinations of parameter fits, so to achieve the best F1-Score (macro weighted) possible. To further reduce the bias due to the random split of the dataset, we repeated the evaluation three times on three different partitions. \changed{Each best classifier is then trained on 75\% of the dataset and tested on the remaining 25\%.} Considering low and high engagement based on 0.75 and 0.25 quantiles implies having heavily unbalanced classes that make the learning process harder. Therefore, we also introduce as tuning parameters the use of well-known under-sampling and over-sampling techniques, i.e., SMOTE and Tomek links~\cite{zeng2016effective, Imbalancedlearn}.

\subsubsection{Results}
\begin{figure}[!h]
\centering
\includegraphics[width=\linewidth]{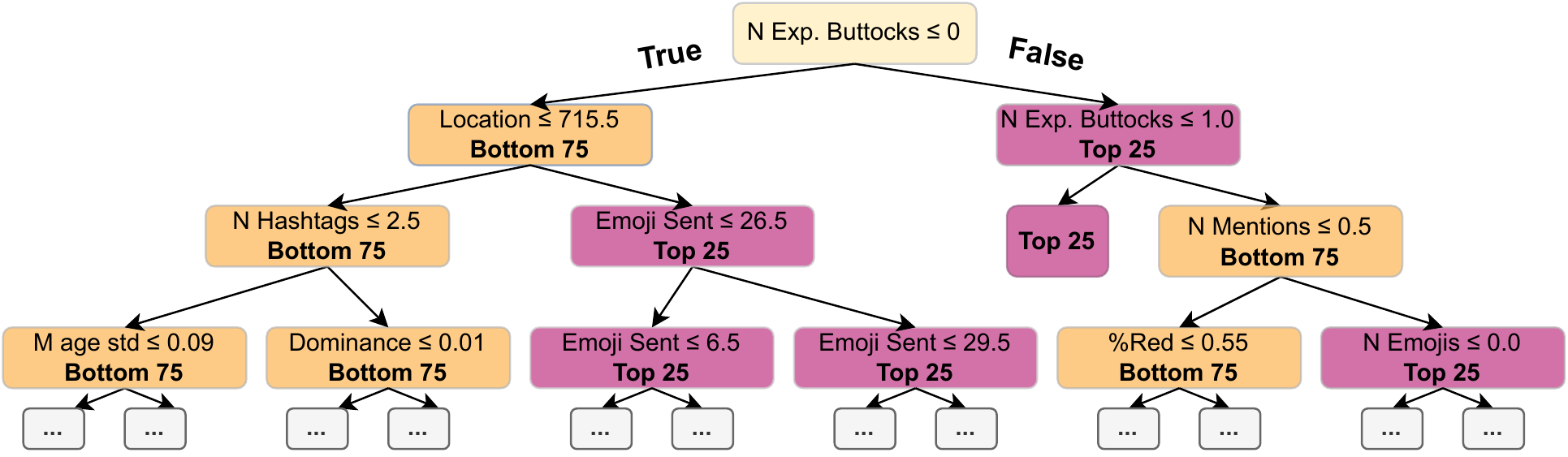}
\caption{Example of guidelines generated by the decision tree for category \feature{Beauty}, tier nano, likes engagement. The representation is limited at a maximum depth of 3.}
\label{fig:decision_tree_example}
\end{figure}

The results on the test sets are reported in Table \ref{tab:all_results_table}. 
\changed{All the results surpass the dummy classifier, showing our method can effectively predict posts' engagement. Moreover, the standard deviations are fairly low, suggesting the models are stable. In terms of Likes, predictions are generally more accurate for Macro and Mega influencers, raging around 60-80\% F1-score (20-40\% better than the dummy). The reason can be that these high-tier influencers tend to be more diversified as we found in the correlation analysis, making some characteristics more effective.   
Accordingly, our classifier exhibited difficulties in the lower tiers of \feature{Fashion}, in which influencers tend to post similar content, and \feature{Other}, in which the content was extremely diverse. On average, we reached the best performances for \feature{Pet}, \feature{Interior}, and \feature{Beauty}. Regarding Comments, we find a behavior similar to Likes, except for the best performances for \feature{Fashion} and \feature{Family}, which appear for Nano influencers. A possible reason is that many Nano influencers might not know the best practices for creating engaging captions, which are strongly correlated to comments engagement as shown in correlation analysis. The best categories we predicted are \feature{Pet}, \feature{Food}, and \feature{Travel}.
Last, we reached the best Likes and Comments prediction score (94\% and 87\%, respectively) for the \feature{Pet} Macro posts.} An example\footnote{All the results are available in our repository.} of guideline with a DT structures is depicted in Figure \ref{fig:decision_tree_example}. Following the nodes conditions (i.e., post characteristics), a label will be assigned when reaching a leaf (i.e., bottom 0.75 or top 0.25 quantile).  We will present more examples of guidelines in §\ref{sec:guidelines}.

\subsubsection{Baselines Comparisons}

As Mazloom et al.~\cite{mazloom2018category}, Gayberi and Oguducu~\cite{gayberi2019popularity}, and other similar studies mentioned, comparison with other works in this area is not completely possible.
The main reasons are the use of private algorithms and data, and how the problem is formulated. Unfortunately, IG policies\footnote{\url{https://help.instagram.com/581066165581870}, accessed: Sep 2022.} never allowed automatic collection and release of common users' posts, forcing previous works to create a new (private) dataset everytime~\cite{mazloom2016multimodal,mazloom2018category,zohourian2018popularity,gayberi2019popularity,de2017predicting,carta2020popularity,purba2021instagram}. Moreover, given the lack of a common dataset to work on, some works focused on a regression problems~\cite{zohourian2018popularity,gayberi2019popularity,purba2021instagram}, other on a classification problem~\cite{zohourian2018popularity, de2017predicting,carta2020popularity}, adopting different metrics, such as the log-normalized number of likes~\cite{gayberi2019popularity, ding2019intrinsic} or the likes divided by the number of followers~\cite{zohourian2018popularity, riis2020limits,purba2021instagram}. \changed{Thus, to set up baselines despite the aforementioned limitations,\footnote{Note that some features used in previous works were not available in our dataset, limiting the comparison.}
we adopted four models: 
(i) I$^2$PA of Ding et al.~\cite{ding2019intrinsic} (the only one publicly released);
(ii) a Decision Neural Network (Dec-NN) for prior works which first extract generic visual or textual features, and then trained a non-interpretable classifier (similar to~\cite{mazloom2018category, gayberi2019popularity, riis2020limits});
(iii) End-to-End Deep NN (EE-DNN) for prior works that relied on end-to-end black-box DL models, giving in input both posts' images and captions simultaneously (similar to~\cite{zhang2018become});
(iv) a stratified dummy classifier,\footnote{\url{https://scikit-learn.org/stable/modules/generated/sklearn.dummy.DummyClassifier.html}}, which predicts targets based on the training set distribution. 
Both Dec-NN and EE-DNN extract posts' image and caption embeddings (through ResNet50~\cite{he2016deep} and Sentence-Bert~\cite{reimers2019sentence}); however, EE-DNN fine-tunes them before the fusion, while Dec-NN receives their early-fusion as input.
The decision is taken through three ReLU feed-forward layers (sizes = 2048+768 $\rightarrow$ 256 $\rightarrow$ 128 $\rightarrow$ 2). 
Both NN were Adam optimized and trained for 50 epochs with early stopping (patience = 5).


The results of Table~\ref{tab:dt_imgpop_dummy} show that our approach outperformed the baselines for each category, except for \feature{Fashion} and \feature{Other}, in which we achieved comparable performance, demonstrating the superiority of our simple DT over Deep Learning models. 
For the categories \feature{Beauty}, \feature{Fitness}, \feature{Food}, \feature{Interior}, \feature{Pet}, and \feature{Travel}, our results are statistically significantly higher than the second-best model (calculated through unpaired \textit{t}-test, two-tailed p-value < 0.05).
Particularly noteworthy is the result against I$^2$PA and EE-DNN, which represents SotA end-to-end DL models. In particular, EE-DNN performs pretty poorly likely because fine-tuning the feature extraction modules led to overfitting. On the other hand, Dec-NN, which is more similar to our strategy, generalized better by not tuning the image and text general representations. Probably, we surpassed such baselines mainly because of the category-related features we extracted, again stressing that developing a cross-category engagement predictor could be unfeasible. Accordingly, we probably could not beat Dec-NN in the \feature{Other} category because of the lack of category-related features.

Although the comparison with previous work is not completely fair for the reasons mentioned above, our results are comparable~\cite{zhang2018become,de2017predicting} or better~\cite{riis2020limits,carta2020popularity,purba2021instagram} than the ones reported on their own data. Anyhow, we remind the reader that our goal is to \textit{explain} the engagement, and not necessarily surpass the prediction of existing non-interpretable models. Last, our dataset was collected using IG APIs from business accounts and is thus shareable. We believe our dataset could serve as a baseline for future works.
}

\begin{table}[h]
\caption{Comparison of Mean F1-Score between our model (DT) and baselines in predicting Likes.}
\label{tab:dt_imgpop_dummy}
\resizebox{.9\columnwidth}{!}{
\begin{threeparttable}[t]
\begin{tabular}{lccccc}
\toprule
\multicolumn{1}{c}{\textbf{Category}} & \textbf{DT (Our)} & \textbf{I$^2$PA} & \textbf{Dec-NN} & \textbf{EE-DNN} & \textbf{Dummy} \\
\midrule
\textbf{Beauty} &  \underline{\bestres{0.65}{0.055}} & \res{0.587}{0.026} & \res{0.582}{0.097} & \res{0.362}{0.037} &\res{0.466}{0.024}\\
\textbf{Fashion} & \res{0.563}{0.030} & \bestres{0.581}{0.019} & \res{0.572}{0.043} & \res{0.327}{0.013} &\res{0.464}{0.008}\\
\textbf{Family} & \bestres{0.568}{0.026} & \res{0.567}{0.019} & \res{0.507}{0.071} & \res{0.347}{0.042} &\res{0.476}{0.017}\\
\textbf{Fitness} & \underline{\bestres{0.640}{0.043}} & \res{0.545}{0.026} & \res{0.511}{0.069} & \res{0.377}{0.062} &\res{0.478}{0.018}\\
\textbf{Food} & \underline{\bestres{0.638}{0.077}} & \res{0.550}{0.030} & \res{0.518}{0.053} & \res{0.464}{0.269} &\res{0.48}{0.031}\\
\textbf{Interior} & \underline{\bestres{0.660}{0.087}} & \res{0.534}{0.056} & \res{0.461}{0.047} & \res{0.309}{0.031} &\res{0.468}{0.022}\\
\textbf{Other} & \res{0.590}{0.026} & \res{0.540}{0.022} & \bestres{0.602}{0.021} & \res{0.318}{0.013} &\res{0.463}{0.004}\\
\textbf{Pet} & \underline{\bestres{0.724}{0.126}} & \res{0.564}{0.046} & \res{0.630}{0.127} & \res{0.342}{0.0727} &\res{0.461}{0.015}\\
\textbf{Travel} & \underline{\bestres{0.642}{0.064}} & \res{0.570}{0.012} & \res{0.473}{0.044} & \res{0.342}{0.069} &\res{0.457}{0.016}\\
\bottomrule
\end{tabular}
\begin{tablenotes}
\item[]\underline{Underlined results} are statistically significantly higher (two-tailed\\ $p$-value < 0.05) than the second-best.
\end{tablenotes}
\end{threeparttable}
}
\end{table}

\subsubsection{Feature Importance}
\changed{Guidelines to create engaging posts results from following the tree generated by the DT classifier. In addition, similarly to correlation analysis, the content creator can inspect the model's feature importance to determine which features are impacting the engagement predictions. Thus, we studied the features used by the models, checking whether they match with correlation results.}
A representative example\footnote{All the results are available in our repository.} of this analysis is shown in Table~\ref{tab:featImp_beauty_micro}, which suggests good correspondence with the factors expressed in §\ref{subsec:correlation}. For example, the presence of common features in small tiers, followed by category-specific features with increasing tier size.
As for the correlation analysis, the number of mentions and whether a location is set result in importance that is inversely proportional to the tier size.

\begin{table}
\caption{Features importance of category Fashion, tier micro.}
    \label{tab:featImp_beauty_micro}
\begin{subtable}[c]{0.45\columnwidth}
\resizebox{.95\columnwidth}{!}{
\begin{tabular}{clcc}
    \toprule
    \multicolumn{1}{c}{\textbf{\#}} & \multicolumn{1}{c}{\textbf{Feature}} & \textbf{Imp.} \\ 
    \midrule
    \textbf{1} & N Mentions & 1.0\\
    \textbf{2} & Age avg & 0.80\\
    \textbf{3} & Dominance & 0.73\\
    \textbf{4} & N Exp.Buttocks & 0.46\\
    \textbf{5} & Outdoor Natural Env. & 0.19\\
    \bottomrule
    \end{tabular}%
    }
\subcaption{Likes.}
\end{subtable}
\begin{subtable}[c]{0.45\columnwidth}
\resizebox{.85\columnwidth}{!}{
\begin{tabular}{clcc}
    \toprule
    \multicolumn{1}{c}{\textbf{\#}} & \multicolumn{1}{c}{\textbf{Feature}} & \textbf{Imp.} \\ 
    \midrule
    \textbf{1} & N Exp. Buttocks & 1.0\\
    \textbf{2} & N Mentions & 0.64\\
    \textbf{3} & Caption Len. & 0.18\\
    \textbf{4} & N Emojis & 0.16\\
    \textbf{5} & Cap. Sent. Magn. & 0.13\\
    \bottomrule
    \end{tabular}%
    }
 
\subcaption{Comments.}
\end{subtable}
\end{table}

\begin{formal}
\textbf{Takeaway:} \textit{A simple and interpretable Decision Tree can outperform Deep Learning algorithms if leveraging domain-knowledge features. Prediction results and feature importance analysis confirm the consideration drawn by feature correlation, showing how similar and dissimilar tiers and category behaves. }
\end{formal}

%% file: Sections/result_table.tex
\begin{table*}[h]
\centering
\caption{Performance of Decision Trees (DT) against a dummy classifier (Dum.). In bold, the best scores for likes and comments for each category. Values reported are F1-Score, macro-weighted \res{mean}{std}.}
\label{tab:all_results_table}
\resizebox{1.0\textwidth}{!}{
\begin{tabular}{lcccc|cccc|cccc|cccc|cccc}
\toprule
 &
  \multicolumn{4}{c|}{\textbf{Nano}} &
  \multicolumn{4}{c|}{\textbf{Micro}} &
  \multicolumn{4}{c|}{\textbf{Mid}} &
  \multicolumn{4}{c|}{\textbf{Macro}} &
  \multicolumn{4}{c}{\textbf{Mega}} \\ 
\multicolumn{1}{l}{} &
  \multicolumn{2}{c}{\textbf{Like}} &
  \multicolumn{2}{c|}{\textbf{Comments}} &
  \multicolumn{2}{c}{\textbf{Like}} &
  \multicolumn{2}{c|}{\textbf{Comments}} &
  \multicolumn{2}{c}{\textbf{Like}} &
  \multicolumn{2}{c|}{\textbf{Comments}} &
  \multicolumn{2}{c}{\textbf{Like}} &
  \multicolumn{2}{c|}{\textbf{Comments}} &
  \multicolumn{2}{c}{\textbf{Like}} &
  \multicolumn{2}{c}{\textbf{Comments}} \\
\multicolumn{1}{l}{} &
  \textbf{DT} &
  \textbf{Dum.} &
  \textbf{DT} &
  \textbf{Dum.} &
  \textbf{DT} &
  \multicolumn{1}{c}{\textbf{Dum.}} &
  \textbf{DT} &
  \textbf{Dum.} &
  \textbf{DT} &
  \multicolumn{1}{c}{\textbf{Dum.}} &
  \textbf{DT} &
  \textbf{Dum.} &
  \textbf{DT} &
  \multicolumn{1}{c}{\textbf{Dum.}} &
  \textbf{DT} &
  \textbf{Dum.} &
  \textbf{DT} &
  \multicolumn{1}{c}{\textbf{Dum.}} &
  \textbf{DT} &
  \multicolumn{1}{c}{\textbf{Dum.}} \\\midrule 
\textbf{Beauty} & \res{0.61}{0.002} & \multicolumn{1}{c}{\res{0.46}{0.004}} & \res{0.65}{0.003} & \res{0.46}{0.004} & 
\res{0.60}{0.003} & \multicolumn{1}{c}{\res{0.47}{0.012}} & \res{0.62}{0.006} & \res{0.46}{0.017} & \res{0.61}{0.004} & \res{0.47}{0.011} & \res{0.61}{0.002} & \res{0.47}{0.016} & \res{0.69}{0.012} & \res{0.45}{0.054} & \res{0.67}{0.011} & \res{0.47}{0.046} & \bestres{0.74}{0.006} & \res{0.48}{0.029} & \bestres{0.68}{0.009} & \res{0.49}{0.025} \\ 
\textbf{Fashion} & \res{0.57}{0.006} & \res{0.47}{0.008} & \bestres{0.62}{0.003} & \res{0.47}{0.001} & 
\res{0.53}{0.002} & \res{0.46}{0.002} & \res{0.57}{0.004} & \res{0.46}{0.004} & \res{0.53}{0.004} & \res{0.47}{0.004} & \res{0.56}{0.007} & \res{0.47}{0.002} & \bestres{0.61}{0.001} & \res{0.47}{0.009} & \res{0.60}{0.011} & \res{0.48}{0.015} & \res{0.57}{0.013} & \res{0.46}{0.011} & \res{0.56}{0.002} & \res{0.47}{0.005} \\ 
\textbf{Family} & \res{0.57}{0.005} & \res{0.47}{0.010} & \bestres{0.59}{0.005} & \res{0.47}{0.007} & 
\res{0.53}{0.002} & \res{0.46}{0.005} & \res{0.56}{0.002} & \res{0.47}{0.008} & \res{0.55}{0.007} & \res{0.48}{0.005} & \res{0.55}{0.002} & \res{0.47}{0.003} & \bestres{0.60}{0.014} & \res{0.47}{0.031} & \bestres{0.59}{0.013} & \res{0.47}{0.017} & \res{0.59}{0.006} & \res{0.47}{0.026} & \res{0.57}{0.001} & \res{0.46}{0.007} \\ 
\textbf{Fitness} & \res{0.62}{0.004} & \res{0.48}{0.005} & \res{0.63}{0.004} & \res{0.47}{0.009} & 
\res{0.60}{0.006} & \res{0.47}{0.009} & \res{0.59}{0.012} & \res{0.47}{0.006} & \res{0.61}{0.010} & \res{0.47}{0.009} & \res{0.61}{0.003} & \res{0.47}{0.010} & \bestres{0.72}{0.010} & \res{0.45}{0.045} & \bestres{0.69}{0.032} & \res{0.46}{0.023} & \res{0.65}{0.013} & \res{0.49}{0.016} & \res{0.61}{0.033} & \res{0.48}{0.010} \\ 
\textbf{Food} & \res{0.59}{0.003} & \res{0.46}{0.003} & \res{0.62}{0.001} & \res{0.47}{0.010} & 
\res{0.57}{0.001} & \res{0.47}{0.007} & \res{0.61}{0.001} & \res{0.47}{0.006} & \res{0.57}{0.005} & \res{0.46}{0.010} & \res{0.63}{0.002} & \res{0.46}{0.011} & \bestres{0.75}{0.008} & \res{0.46}{0.020} & \res{0.71}{0.024} & \res{0.47}{0.061} & \res{0.71}{0.023} & \res{0.57}{0.091} & \bestres{0.72}{0.029} & \res{0.53}{0.098} \\ 
\textbf{Interior} & \res{0.59}{0.002} & \res{0.48}{0.012} & \res{0.62}{0.003} & \res{0.46}{0.01} & 
\res{0.59}{0.003} & \res{0.47}{0.002} & \res{0.63}{0.001} & \res{0.47}{0.006} & \res{0.58}{0.010} & \res{0.45}{0.012} & \res{0.63}{0.003} & \res{0.47}{0.011} & \res{0.76}{0.033} & \res{0.48}{0.021} & \bestres{0.83}{0.020} & \res{0.49}{0.015} & \bestres{0.77}{0.011} & \res{0.44}{0.083} & \res{0.74}{0.027} & \res{0.45}{0.074} \\ 
\textbf{Other} & \res{0.58}{0.002} & \res{0.47}{0.001} & \res{0.57}{0.0001} & \res{0.47}{0.004} & 
\res{0.55}{0.001} & \res{0.47}{0.0021} & \res{0.56}{0.005} & \res{0.47}{0.002} & \res{0.59}{0.001} & \res{0.47}{0.006} & \res{0.58}{0.001} & \res{0.46}{0.005} & \bestres{0.63}{0.003} & \res{0.46}{0.013} & \bestres{0.59}{0.008} & \res{0.47}{0.023} & \res{0.60}{0.003} & \res{0.46}{0.005} & \res{0.58}{0.008} & \res{0.46}{0.012} \\ 
\textbf{Pet} & \res{0.69}{0.004} & \res{0.47}{0.019} & \res{0.72}{0.006} & \res{0.49}{0.009} & 
\res{0.60}{0.008} & \res{0.45}{0.018} & \res{0.62}{0.005} & \res{0.46}{0.007} & \res{0.61}{0.018} & \res{0.46}{0.006} & \res{0.64}{0.003} & \res{0.47}{0.024} & \bestres{0.94}{0.057} & \res{0.42}{0.042} & \bestres{0.87}{0.048} & \res{0.42}{0.042} & \res{0.78}{0.045} & \res{0.51}{0.032} & \res{0.77}{0.012} & \res{0.5}{0.065} \\ 
\textbf{Travel} & \res{0.60}{0.001} & \res{0.47}{0.0042} & \res{0.61}{0.001} & \res{0.47}{0.006} & 
\res{0.59}{0.0012} & \res{0.47}{0.003} & \res{0.63}{0.002} & \res{0.47}{0.007} & \res{0.58}{0.004} & \res{0.47}{0.006} & \res{0.64}{0.009} & \res{0.48}{0.010} & \bestres{0.72}{0.008} & \res{0.45}{0.024} & \res{0.67}{0.050} & \res{0.47}{0.012} & \bestres{0.72}{0.010} & \res{0.46}{0.031} & \bestres{0.71}{0.017} & \res{0.42}{0.032} \\\bottomrule

\end{tabular}
}
\end{table*}

%% file: Sections/06-Unsupervised.tex
\section{Spotting Instagram Hot Topics}
\label{sec:unsupervised}
Our features allow us to predict a post's engagement with good accuracy, but there is room for improvement. In our \textit{interpretable} approach, features have to be extracted a priori instead of being learned ``automatically'' by a deep learning model. Thus, our features are limited by our educated guesses of what could be engaging, and by the concepts obtainable through existing SotA deep learning models. For instance, if available, we would have used a love or a marriage scene detector, which are likely to produce high engagement.  
Although such detectors could be implemented through classical approaches (e.g., by fine-tuning an image recognition NN like ResNet~\cite{he2016deep}), we opted for defining an unsupervised strategy to detect \textit{general hot topics}. 
In particular, we aim to find (if any) topics or concepts that, if present in a post, would create high engagement independently from the publisher. \changed{In this context, unsupervised means we make no assumptions on which topics are engaging (as we did to extract category-related features for §\ref{sec:prediction}), but rather explore users' interests~\cite{zarrinkalam2020extracting}.}

From §\ref{sec:prediction} we learned that likes and comments are mainly driven by the image and caption, respectively. Thus, in the next experiments, we focus on finding likes-related hot topics through visual features, and comments-related hot topics through textual features. We now present our methodology and findings.

\subsection{Methodology}
\label{subsec:pure}
The idea behind our method is to group together semantically similar images and captions, and observe whether some of these groups reach high engagement on average. 

\textbf{Embeddings.} To define image and text semantic similarity, we rely on the concept of embedding. 
An embedding is a vector representation of an object (e.g., image, text) in which objects with similar semantics have similar vector profiles~\cite{le2014distributed}. Embeddings are usually extracted by taking the output of the penultimate layer of a deep neural network performing a classification task. In our experiments, we retrieved image embeddings using ResNet50~\cite{he2016deep} pre-trained on the ImageNet dataset, and text-embeddings using Sentence-Bert~\cite{reimers2019sentence} (in particular, in its version \textit{all-mpnet-base-v2}~\cite{mpnet}). Before extracting the text embeddings, we translated non-English text leveraging Google Cloud Platform ~\cite{gcp}, so to perform language detection and translation automatically.

\begin{figure}[!ht]
    \centering
    \subfloat[\centering Likes]
    {{\includegraphics[width=0.6\linewidth]{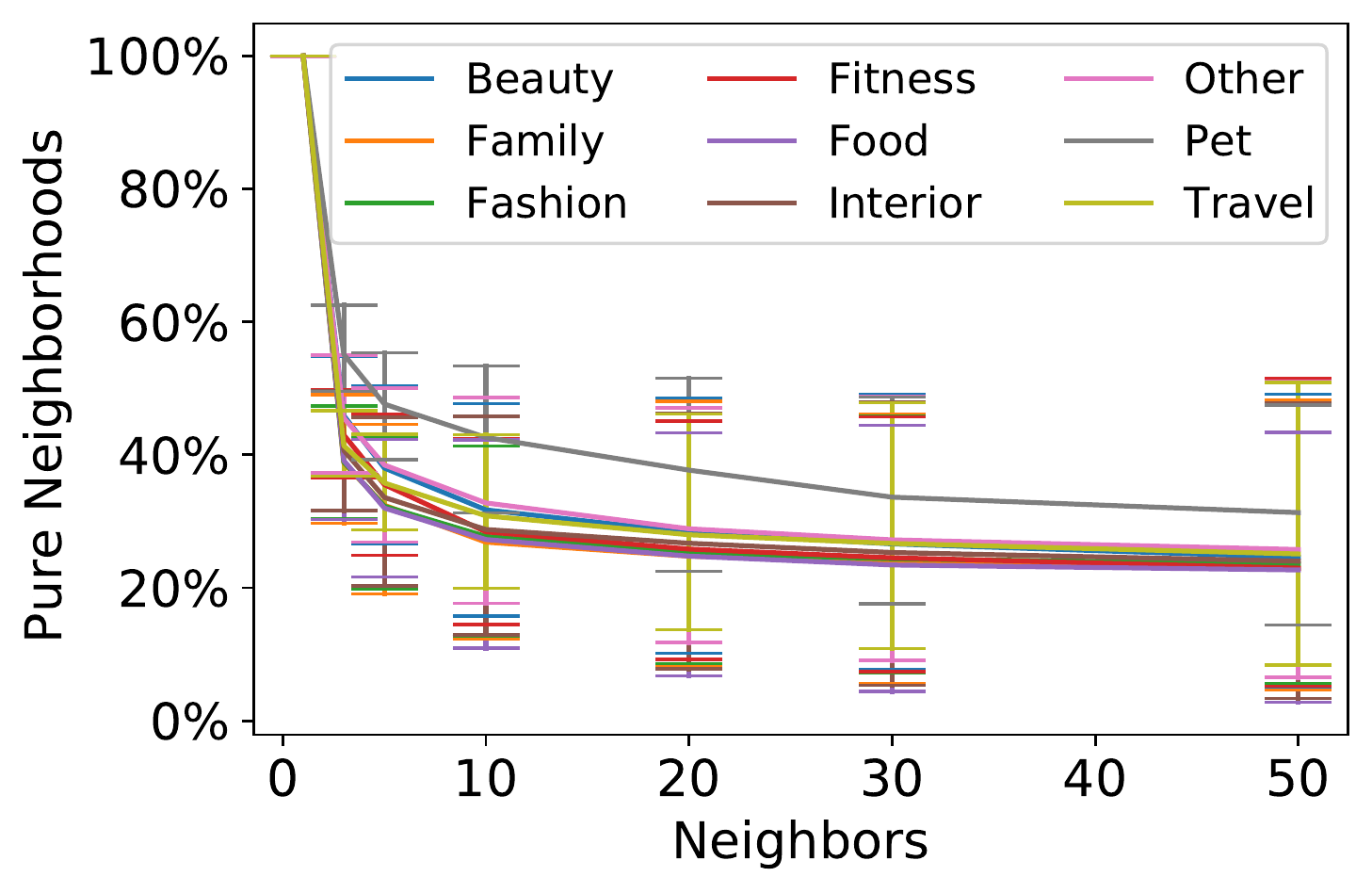} }}%
    \hspace{-4pt}
    \subfloat[\centering Comments]
    {{\includegraphics[width=0.6\linewidth]{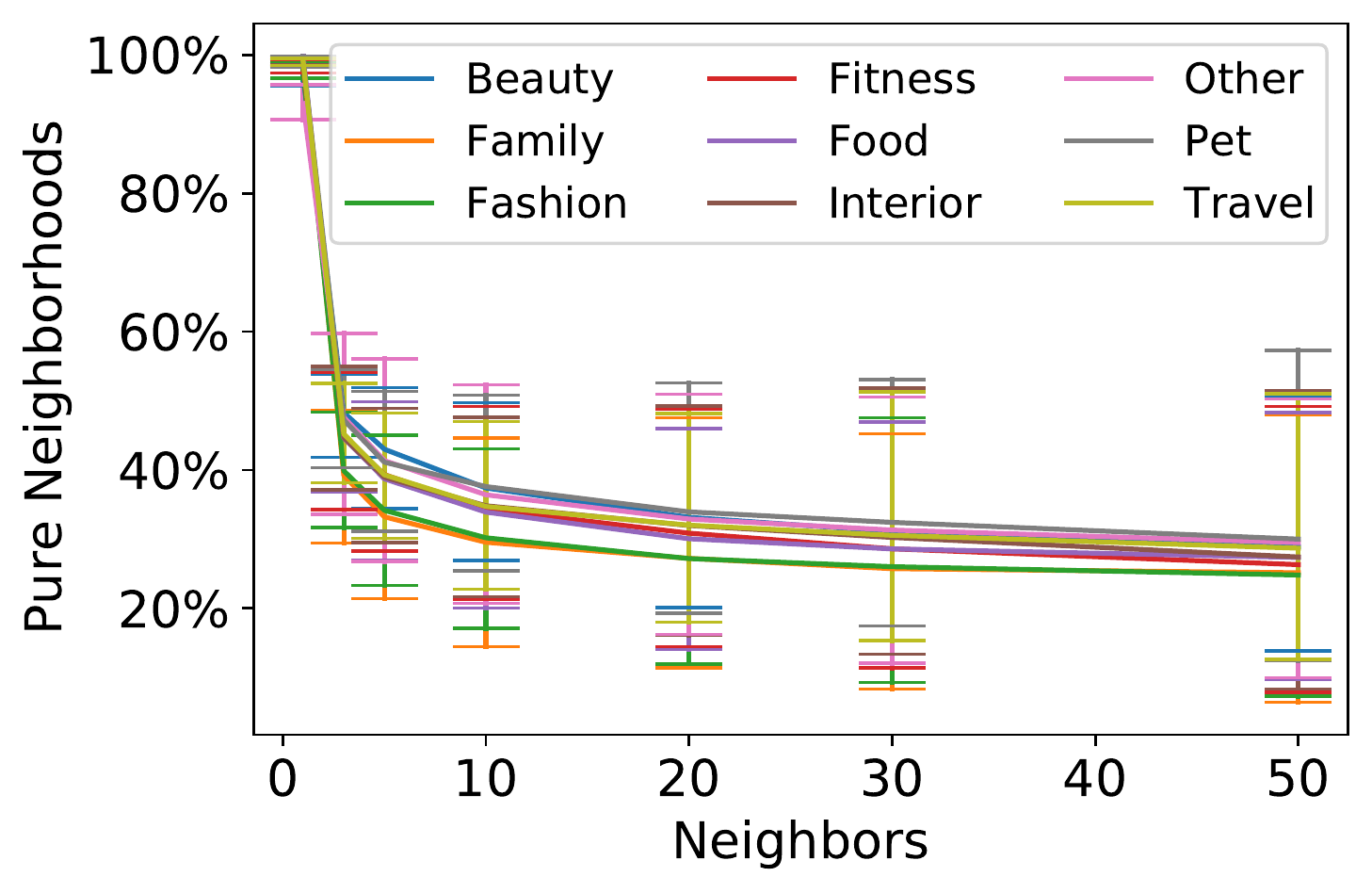} }}%
    \caption{Percentage of pure neighborhoods in mid-tier for engagement metrics.}%
    \label{fig:pure}%
\end{figure}

\textbf{Semantically Similar Neighborhood.}
As a first approach, we could create clusters of similar images or captions, and see whether some clusters present higher engagement than others. However, as shown in the literature~\cite{niu2021spice}, current cluster algorithms suffer the decision of the number of clusters beforehand. Moreover, finding hot topics is challenging~\cite{zhang2018trending}, since they could be small and lost in a big cluster. Thus, we prefer a Nearest Neighbors approach, to find neighborhoods of points with similar engagement. In particular, we first divided our posts into five engagement classes determined by the percentiles [0-20, 20-40, 40-60, 60-80, 80-100], saving the thresholds of each percentile. Then, for each point, we search its N nearest neighbors, calculate their engagement average, and see whether the average falls in the same engagement class as the point under consideration. If so, that neighborhood is considered ``pure'', and new posts falling in it would likely produce that particular class of engagement. 
To find the nearest neighbors, we first reduced the dimensionality of the embeddings using PCA (100 components), and then applied Nearest Neighbor algorithm leveraging Scikit-Learn implementation \cite{sklearn} using Euclidean Distance as the distance metric.  
Figure~\ref{fig:pure} depicts the percentage of pure neighborhoods for different N = [1, 3, 5, 10, 20, 30, 50] for the mid-tier. On average, around 20\% (N=50) of the points are in a pure neighborhood, which suggests that some topics are more (or less) engaging than others. The pet category presents the highest average, probably because its topics can be the species and breed of animals (visually similar), and some could be liked more (or less) than others.

\subsection{General vs User-specific Hot topics}\label{subsec:hottopics}
\begin{figure}[!h]
\hspace{-1cm}
\includegraphics[width=0.5\linewidth]{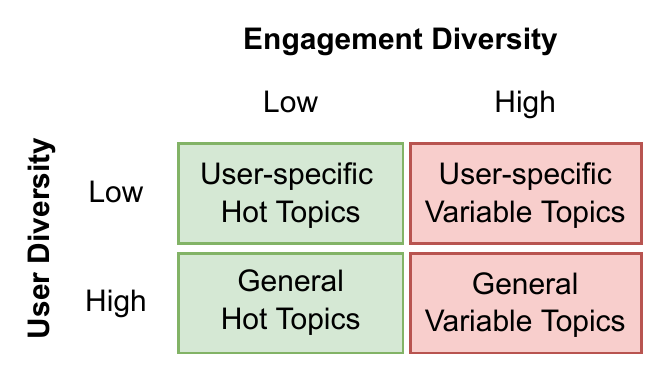}
\caption{Types of topics for Engagement and User Diversity.}
\label{fig:conf}
\end{figure}

The percentage of the pure neighborhoods found in §\ref{subsec:pure} is comprehensive of neighborhoods made only by a single influencer, i.e., not a general hot topic. In this case, we identified what we can call a user-specific hot topic, which is very useful to understand what topic is engaging (or not) for that particular influencer. Thus, what differentiates general vs user-specific hot topics is how many influencers participate in a pure neighborhood, and with how many posts. We call this parameter \textit{User Diversity}. To calculate it, we took inspiration from the Simpson's Diversity Index~\cite{simpson1949measurement}, used in ecology to quantify the biodiversity of a habitat. It takes into account the number of species present, as well as the abundance of each species. The diversity index $D$ is expressed as:

\begin{equation}
    D = 1 - \frac{\sum_{i = 1}^{K}n_{i}(n_{i}-1)}{N(N-1)},
\end{equation}
\noindent
where $N$ is the total sample size, $K$ the number of species, and $n_{i}$ is the number of organisms of the $i^{th}$ specie. $D$ ranges from 0 to 1, where 0 is minimum diversity and 1 is maximum diversity. In our scenario, the species are the influencers, and the organisms the posts. 
Similarly, we can define an \textit{Engagement Diversity}, which measures the posts' diversity in terms of engagement. This metric is needed since we created pure neighborhoods based only the average engagement of their posts, therefore, within a pure neighborhood there can be posts belonging to different engagement classes. To recap, by measuring the \textit{Engagement} and \textit{User Diversity} of our pure neighborhoods, we can define topics as depicted in Figure~\ref{fig:conf}. Obviously, we are more interested in the green part, since neighborhoods with high engagement diversity are less reliable. We set the threshold between low and high at 0.5.

\begin{figure*}
    \centering
    \includegraphics[width=0.98\linewidth]{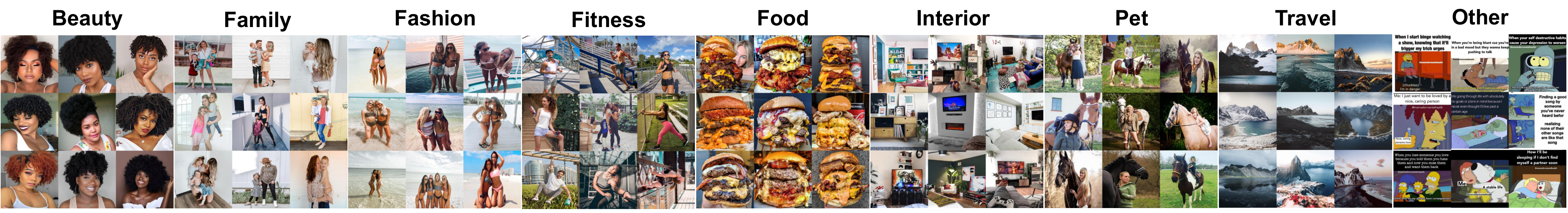}
    \caption{Examples of hot topics found in our categories.}
    \label{fig:hot_topics}
\end{figure*}

\textbf{Findings.} We concentrated our research of hot topics only on the highest engagement class (i.e., posts falling in top 20\% percentile), focusing on each tier and category differently. During the automatic search, we removed neighborhoods that overlapped for more than 80\% of the points. We explored the resulting neighborhoods and found several hot topics, which we did not think about in the feature extraction phase, and could not be detected by SotA models, confirming the benefits of this unsupervised research. All the neighborhoods can be browsed on our repository, while in Figure~\ref{fig:hot_topics} we reported an example for each category. For instance, we found ``mother with her child'' for \feature{Family}, ``two (or more) girls in bikini'' for \feature{Fashion}, or ``girl/kid near/riding a horse'' for \feature{Pet}.  
For captions, we found less category-specific hot topics, and more common strategies. For instance, giveaways attracts a lot of comments, since participants usually have to comment and tag other friends. Furthermore, we found the a working strategy of asking users' opinions on a topic, general questions, or requests for upcoming content.
More general hot topics are presented in §\ref{sec:guidelines}.

\begin{formal}
\textbf{Takeaway:} \textit{Instagram offers both visual and textual hot topics that are likely to generate high engagement levels. Captions tend to use similar strategies across categories despite visual hot topics being category-specific.}
\end{formal}

%% file: Sections/07-Guidelines.tex
\section{Guidelines Insights}
\label{sec:guidelines}
This section provides some guidelines to get more Likes and Comments for each category, resulting from both DT engagement classifiers (§\ref{sec:prediction}) and hot topics detection (§\ref{sec:unsupervised}). We also provide some suggestions to make an engaging caption.

\subsection{Guidelines for Likes, Comments, and Topics}
Each category presents different characteristics to build engagement. We now present guidelines to get more Likes and Comments along the hot topics we found.  

\textbf{Beauty.} \changed{Likes are mainly driven by exposed buttocks and feet, a high image pleasure, and positive emoji sentiment. Exposed buttocks also generate many comments, as well as a low age average, having the location set, and the use of many mentions. Wavy hair is much appreciated, and hot topics include couples and eye make-up with perfect eyebrows. Users usually love when the influencers receive new make-up products, recreate famous make-up (e.g., from movies), and talk about personal problems.

\textbf{Family.} Likes and Comments are driven by similar factors. People's features like age and gender are predominant. The mean age of female subjects should be low, with a high standard deviation. This suggests that mothers with children are a hot topic, as detected in §\ref{subsec:hottopics}. Indoor or outdoor-natural environments are preferred, and location, colors, and the number of mentions are highly impacting. As hot topics, we found pregnancy, childbirth, and body changes, during which followers feel closer to the influencer.

\textbf{Fashion.} In terms of Likes, a higher number of mentions is suggested for small tiers, whereas colors-related features (e.g., dominance, arousal) contribute heavily to high-tier influencers. A predominant role is held by exposed buttocks, which contribute to both likes and comments. Exposed feet generate many comments, as well as outdoor pictures, short captions with many hashtags, and positive emoji sentiments. As hot topics, girls in bikinis and men with six-packs are successful. Discussing outfits for special events is highly engaging, such as traveling, going to concerts, birthdays, gallant dinners, or simply starting the week.  

\textbf{Fitness.} Likes are driven by warm colors, and high dominance of female subjects with low age standard deviation, preferably in their workout outfit. A short caption with positive sentiment helps in receiving both likes and comments. Low arousal generates many comments, according to the body transformation hot topic. The caption should motivate people to try harder in their workouts. 

\textbf{Food.} Males are more common and generate more likes in this category. Extreme burgers and spirits are highly appreciated, as well as perfect and very colored food compositions. Pictures in kitchens or outside restaurants help to get likes. The location is important for getting comments, as well as high arousal and a positive caption with many Emojis. The caption should include a brief description of the plate and questions about favorite food. Pizza days, chocolate, and vegan food are often in the middle of heated discussions.

\textbf{Interior.} To get likes, indoor environments like a living room and cold colors are preferred, as well as the presence of kids and female subjects. The location is relevant for both likes and comments, but avoid commercial buildings and food pictures. Luminous and pleasant pictures generate more comments, as well as the presence of animals.
A good caption combines positive Emoji sentiment, a few hashtags, and general questions, like what to do on the weekend.

\textbf{Pet.} Pictures should be in an indoor or outdoor-natural environment, use warm colors, and convey a positive sentiment to get many likes. The use of the location and mentions helps a lot for comments, as well as very high or very low animal cuteness. Among the most loved animals, we found horses, exotic animals, Siamese cats, and dogs with clothes and ribbons. Many comments will arrive along with a new family member!

\textbf{Travel.} Likes are gained primarily by female subjects and a low number of male presence, with a generally low age standard deviation and a low minimum age. This suggests that travel pictures of young friends or groups of the same age are highly engaging. To get many comments, besides the importance of outdoor-related features, the sentiment conveyed by the text should tend to be positive. Further, hashtags and mentions are crucial, and the picture should be pleasant and arousing, with low dominance. Pictures near the sea are highly appreciated, and users' engage more with summer and holiday posts.

\textbf{Other.} More likes are obtained by females in indoor places with cold colors and low dominance. Many emojis and short captions help too. Men of the same age in a single picture get many comments, using a neutral sentiment in the caption. Hot topics are many, for example, football, memes, or superheroes. The captions tend to be funny and quite short.

}

\subsection{Guidelines for Engaging Captions}
Even if each category and tier require a specific caption to build engagement, we identified some common strategies to generate highly engaging ones. From our explorations, we identified a typical pattern among most hot captions, i.e., asking questions to the audience. Such questions can be very generic (e.g., ``how do you feel today?''), or topic-specific (e.g., ``which outfit do you prefer?''), which helps engage the users. Moreover, creators often ask people to perform particular actions, such as tagging or sharing content with friends. This behavior, known as \textit{call to action}, usually generates a lot of engagement. Last, hashtags are generally at the end of the caption, often separated by the rest of the text with one point or dash per line. This behavior forces the users to click on the ``View more'' button to see the whole caption, generating more engagement.

\subsection{Limitations}
\changed{
Our guidelines are the result of analyzing the biggest IG dataset ever released, composed of around 10M posts created by 34K influencers. Nevertheless, it does not include many categories of interest (e.g., sport, cinema, music), and what people like as well as hot topics could change over time. 
However, we presented two methodologies (supervised §\ref{sec:prediction} and unsupervised §\ref{sec:unsupervised}) that can be applied to any category (possibly enhancing the feature set) at any time, by taking a ``snapshot''
of IG content produced by influencers of a target category and tier.
Moreover, a reader might be concerned that the IG algorithm started considering comments to recommend posts in 2021, whereas our dataset is from 2020. We remark that our paper's aim is to explain which post's features induce users to generate more comments (which now are a crucial factor), and not how IG is now recommending posts to users. Indeed, as for likes, the main reason users leave comments is based on the posts themselves~\cite{aldous2019view}, not whether users see the posts.
}

%% file: Sections/08-Conclusion.tex
\section{Conclusion}
\label{sec:conclusion}
In this work, we aimed to close the gap from previous works, explaining the underlying mechanisms of IG engagement and focusing on the interpretable models. In this way, it is possible to create engaging IG content by design, following predefined guidelines, saving time and money.
Through a careful and all-inclusive process of feature extraction, we trained predictors that achieved up to 94\% of F1-Score. In particular, our results show that likes are mainly driven by images, while comments are primarily stimulated by captions. Further, we demonstrated how influencers' behavior becomes more category-specific as their tier increases. Last, we proposed a novel unsupervised approach for detecting and analyzing hot topics, to better understand the inner dynamics of each category. 

In the future, we plan to improve the predictions through a model that integrates hot topic extraction. Furthermore, more categories should be studied, and a metric that combines likes and comments should be introduced to better understand their relationship. \changed{Regarding these metrics, they could be polished by removing fake engagement, for instance. However, as of now, we have no evidence IG algorithm is accounting for such differences.
Last, geography should be taken into account to understand whether it impacts engagements mechanisms}

%% file: Main.bbl

\begin{thebibliography}{59}


\ifx \showCODEN    \undefined \def \showCODEN     #1{\unskip}     \fi
\ifx \showDOI      \undefined \def \showDOI       #1{#1}\fi
\ifx \showISBNx    \undefined \def \showISBNx     #1{\unskip}     \fi
\ifx \showISBNxiii \undefined \def \showISBNxiii  #1{\unskip}     \fi
\ifx \showISSN     \undefined \def \showISSN      #1{\unskip}     \fi
\ifx \showLCCN     \undefined \def \showLCCN      #1{\unskip}     \fi
\ifx \shownote     \undefined \def \shownote      #1{#1}          \fi
\ifx \showarticletitle \undefined \def \showarticletitle #1{#1}   \fi
\ifx \showURL      \undefined \def \showURL       {\relax}        \fi
\providecommand\bibfield[2]{#2}
\providecommand\bibinfo[2]{#2}
\providecommand\natexlab[1]{#1}
\providecommand\showeprint[2][]{arXiv:#2}

\bibitem[Aldous et~al\mbox{.}(2019)]%
        {aldous2019view}
\bibfield{author}{\bibinfo{person}{Kholoud~Khalil Aldous},
  \bibinfo{person}{Jisun An}, {and} \bibinfo{person}{Bernard~J Jansen}.}
  \bibinfo{year}{2019}\natexlab{}.
\newblock \showarticletitle{View, like, comment, post: Analyzing user
  engagement by topic at 4 levels across 5 social media platforms for 53 news
  organizations}. In \bibinfo{booktitle}{\emph{Proceedings of the International
  AAAI Conference on Web and Social Media}}, Vol.~\bibinfo{volume}{13}.
  \bibinfo{pages}{47--57}.
\newblock


\bibitem[Campos et~al\mbox{.}(2017)]%
        {campos2017pixels}
\bibfield{author}{\bibinfo{person}{Victor Campos}, \bibinfo{person}{Brendan
  Jou}, {and} \bibinfo{person}{Xavier Giro-i Nieto}.}
  \bibinfo{year}{2017}\natexlab{}.
\newblock \showarticletitle{From pixels to sentiment: Fine-tuning CNNs for
  visual sentiment prediction}.
\newblock \bibinfo{journal}{\emph{Image and Vision Computing}}
  \bibinfo{volume}{65} (\bibinfo{year}{2017}), \bibinfo{pages}{15--22}.
\newblock


\bibitem[Carta et~al\mbox{.}(2020)]%
        {carta2020popularity}
\bibfield{author}{\bibinfo{person}{Salvatore Carta},
  \bibinfo{person}{Alessandro~Sebastian Podda},
  \bibinfo{person}{Diego~Reforgiato Recupero}, \bibinfo{person}{Roberto Saia},
  {and} \bibinfo{person}{Giovanni Usai}.} \bibinfo{year}{2020}\natexlab{}.
\newblock \showarticletitle{Popularity prediction of instagram posts}.
\newblock \bibinfo{journal}{\emph{Information}} \bibinfo{volume}{11},
  \bibinfo{number}{9} (\bibinfo{year}{2020}), \bibinfo{pages}{453}.
\newblock


\bibitem[Chandra~Guntuku et~al\mbox{.}(2019)]%
        {twittercolordepression}
\bibfield{author}{\bibinfo{person}{Sharath Chandra~Guntuku},
  \bibinfo{person}{Daniel Preotiuc-Pietro}, \bibinfo{person}{Johannes~C.
  Eichstaedt}, {and} \bibinfo{person}{Lyle~H. Ungar}.}
  \bibinfo{year}{2019}\natexlab{}.
\newblock \showarticletitle{What Twitter Profile and Posted Images Reveal about
  Depression and Anxiety}.
\newblock \bibinfo{journal}{\emph{Proceedings of the International AAAI
  Conference on Web and Social Media}} \bibinfo{volume}{13},
  \bibinfo{number}{01} (\bibinfo{date}{Jul.} \bibinfo{year}{2019}),
  \bibinfo{pages}{236--246}.
\newblock


\bibitem[Conti et~al\mbox{.}(2022)]%
        {conti2022virtual}
\bibfield{author}{\bibinfo{person}{Mauro Conti}, \bibinfo{person}{Jenil
  Gathani}, {and} \bibinfo{person}{Pier~Paolo Tricomi}.}
  \bibinfo{year}{2022}\natexlab{}.
\newblock \showarticletitle{Virtual Influencers in Online Social Media}.
\newblock \bibinfo{journal}{\emph{IEEE Communications Magazine}}
  (\bibinfo{year}{2022}).
\newblock


\bibitem[De et~al\mbox{.}(2017)]%
        {de2017predicting}
\bibfield{author}{\bibinfo{person}{Shaunak De}, \bibinfo{person}{Abhishek
  Maity}, \bibinfo{person}{Vritti Goel}, \bibinfo{person}{Sanjay Shitole},
  {and} \bibinfo{person}{Avik Bhattacharya}.} \bibinfo{year}{2017}\natexlab{}.
\newblock \showarticletitle{Predicting the popularity of instagram posts for a
  lifestyle magazine using deep learning}. In \bibinfo{booktitle}{\emph{2017
  2nd international conference on communication systems, computing and IT
  applications (CSCITA)}}. IEEE, \bibinfo{pages}{174--177}.
\newblock


\bibitem[Deng et~al\mbox{.}(2020)]%
        {retinafacepaper}
\bibfield{author}{\bibinfo{person}{Jiankang Deng}, \bibinfo{person}{Jia Guo},
  \bibinfo{person}{Evangelos Ververas}, \bibinfo{person}{Irene Kotsia}, {and}
  \bibinfo{person}{Stefanos Zafeiriou}.} \bibinfo{year}{2020}\natexlab{}.
\newblock \showarticletitle{Retinaface: Single-shot multi-level face
  localisation in the wild}. In \bibinfo{booktitle}{\emph{Proceedings of the
  IEEE/CVF conference on computer vision and pattern recognition}}.
  \bibinfo{pages}{5203--5212}.
\newblock


\bibitem[Ding et~al\mbox{.}(2019)]%
        {ding2019intrinsic}
\bibfield{author}{\bibinfo{person}{Keyan Ding}, \bibinfo{person}{Kede Ma},
  {and} \bibinfo{person}{Shiqi Wang}.} \bibinfo{year}{2019}\natexlab{}.
\newblock \showarticletitle{Intrinsic image popularity assessment}. In
  \bibinfo{booktitle}{\emph{Proceedings of the 27th ACM International
  Conference on Multimedia}}. \bibinfo{pages}{1979--1987}.
\newblock


\bibitem[Face(2022)]%
        {mpnet}
\bibfield{author}{\bibinfo{person}{Hugging Face}.}
  \bibinfo{year}{2022}\natexlab{}.
\newblock
\newblock
\urldef\tempurl%
\url{https://huggingface.co/sentence-transformers/all-mpnet-base-v2}
\showURL{%
\tempurl}
\newblock
\shownote{Accessed: August 2022}.


\bibitem[Gayberi and Oguducu(2019)]%
        {gayberi2019popularity}
\bibfield{author}{\bibinfo{person}{Mehmetcan Gayberi} {and}
  \bibinfo{person}{Sule~Gunduz Oguducu}.} \bibinfo{year}{2019}\natexlab{}.
\newblock \showarticletitle{Popularity prediction of posts in social networks
  based on user, post and image features}. In
  \bibinfo{booktitle}{\emph{Proceedings of the 11th International Conference on
  Management of Digital EcoSystems}}. \bibinfo{pages}{9--15}.
\newblock


\bibitem[Geyser(2021)]%
        {engagement_what_is_useful}
\bibfield{author}{\bibinfo{person}{Werner Geyser}.}
  \bibinfo{year}{2021}\natexlab{}.
\newblock \bibinfo{title}{The State of Influencer Marketing 2021: Benchmark
  Report}.
\newblock
\newblock
\urldef\tempurl%
\url{https://influencermarketinghub.com/influencer-marketing-benchmark-report-2021/}
\showURL{%
\tempurl}
\newblock
\shownote{Accessed: August 2022}.


\bibitem[Google(2022)]%
        {gcp}
\bibfield{author}{\bibinfo{person}{Google}.} \bibinfo{year}{2022}\natexlab{}.
\newblock \bibinfo{booktitle}{\emph{Google Cloud Platform}}.
\newblock
\urldef\tempurl%
\url{https://cloud.google.com/}
\showURL{%
\tempurl}


\bibitem[He et~al\mbox{.}(2016)]%
        {he2016deep}
\bibfield{author}{\bibinfo{person}{Kaiming He}, \bibinfo{person}{Xiangyu
  Zhang}, \bibinfo{person}{Shaoqing Ren}, {and} \bibinfo{person}{Jian Sun}.}
  \bibinfo{year}{2016}\natexlab{}.
\newblock \showarticletitle{Deep residual learning for image recognition}. In
  \bibinfo{booktitle}{\emph{Proceedings of the IEEE conference on computer
  vision and pattern recognition}}. \bibinfo{pages}{770--778}.
\newblock


\bibitem[Hee et~al\mbox{.}(2022)]%
        {hee2022explaining}
\bibfield{author}{\bibinfo{person}{Ming~Shan Hee}, \bibinfo{person}{Roy Ka-Wei
  Lee}, {and} \bibinfo{person}{Wen-Haw Chong}.}
  \bibinfo{year}{2022}\natexlab{}.
\newblock \showarticletitle{On Explaining Multimodal Hateful Meme Detection
  Models}. In \bibinfo{booktitle}{\emph{Proceedings of the ACM Web Conference
  2022}}. \bibinfo{pages}{3651--3655}.
\newblock


\bibitem[HypeAuditor(2022)]%
        {IGcatdistr}
\bibfield{author}{\bibinfo{person}{HypeAuditor}.}
  \bibinfo{year}{2022}\natexlab{}.
\newblock \bibinfo{title}{State of Influencer Marketing 2022}.
\newblock
\newblock
\urldef\tempurl%
\url{https://hypeauditor.com/blog/wp-content/uploads/2022/01/US-State-of-Influencer-Marketing-2022.pdf}
\showURL{%
\tempurl}
\newblock
\shownote{Accessed: August 2022}.


\bibitem[Insiderintelligence(2021)]%
        {instagramuserstatisticstrends_revenue2021}
\bibfield{author}{\bibinfo{person}{Insiderintelligence}.}
  \bibinfo{year}{2021}\natexlab{}.
\newblock \bibinfo{title}{Instagram will net more US ad revenues than core
  Facebook platform}.
\newblock
\newblock
\urldef\tempurl%
\url{https://www.insiderintelligence.com/content/instagram-will-net-more-us-ad-revenues-than-core-facebook-platform}
\showURL{%
\tempurl}
\newblock
\shownote{Accessed: August 2022}.


\bibitem[Insiderintelligence(2022)]%
        {instagramuserstatisticstrends}
\bibfield{author}{\bibinfo{person}{Insiderintelligence}.}
  \bibinfo{year}{2022}\natexlab{}.
\newblock \bibinfo{title}{Instagram in 2022: Global user statistics,
  demographics and marketing trends to know}.
\newblock
\newblock
\urldef\tempurl%
\url{https://www.insiderintelligence.com/insights/instagram-user-statistics-trends/}
\showURL{%
\tempurl}
\newblock
\shownote{Accessed: August 2022}.


\bibitem[Kangasharju(2022)]%
        {cutedetector}
\bibfield{author}{\bibinfo{person}{Jaakko Kangasharju}.}
  \bibinfo{year}{2022}\natexlab{}.
\newblock \bibinfo{title}{Cuteness Detector}.
\newblock
  \bibinfo{howpublished}{https://github.com/asharov/cute-animal-detector}.
\newblock


\bibitem[Kim et~al\mbox{.}(2020)]%
        {datasetkim2020multimodal}
\bibfield{author}{\bibinfo{person}{Seungbae Kim}, \bibinfo{person}{Jyun-Yu
  Jiang}, \bibinfo{person}{Masaki Nakada}, \bibinfo{person}{Jinyoung Han},
  {and} \bibinfo{person}{Wei Wang}.} \bibinfo{year}{2020}\natexlab{}.
\newblock \showarticletitle{Multimodal Post Attentive Profiling for Influencer
  Marketing}. In \bibinfo{booktitle}{\emph{Proceedings of The Web Conference
  2020}}. \bibinfo{pages}{2878--2884}.
\newblock


\bibitem[Kong et~al\mbox{.}(2016)]%
        {kong2016photo}
\bibfield{author}{\bibinfo{person}{Shu Kong}, \bibinfo{person}{Xiaohui Shen},
  \bibinfo{person}{Zhe Lin}, \bibinfo{person}{Radomir Mech}, {and}
  \bibinfo{person}{Charless Fowlkes}.} \bibinfo{year}{2016}\natexlab{}.
\newblock \showarticletitle{Photo aesthetics ranking network with attributes
  and content adaptation}. In \bibinfo{booktitle}{\emph{European conference on
  computer vision}}. Springer, \bibinfo{pages}{662--679}.
\newblock


\bibitem[Kralj~Novak et~al\mbox{.}(2015)]%
        {10.1371/journal.pone.0144296}
\bibfield{author}{\bibinfo{person}{Petra Kralj~Novak}, \bibinfo{person}{Jasmina
  Smailović}, \bibinfo{person}{Borut Sluban}, {and} \bibinfo{person}{Igor
  Mozetič}.} \bibinfo{year}{2015}\natexlab{}.
\newblock \showarticletitle{Sentiment of Emojis}.
\newblock \bibinfo{journal}{\emph{PLOS ONE}} \bibinfo{volume}{10},
  \bibinfo{number}{12} (\bibinfo{date}{12} \bibinfo{year}{2015}),
  \bibinfo{pages}{1--22}.
\newblock


\bibitem[Le and Mikolov(2014)]%
        {le2014distributed}
\bibfield{author}{\bibinfo{person}{Quoc Le} {and} \bibinfo{person}{Tomas
  Mikolov}.} \bibinfo{year}{2014}\natexlab{}.
\newblock \showarticletitle{Distributed representations of sentences and
  documents}. In \bibinfo{booktitle}{\emph{International conference on machine
  learning}}. PMLR, \bibinfo{pages}{1188--1196}.
\newblock


\bibitem[Lema{{\^i}}tre et~al\mbox{.}(2017)]%
        {Imbalancedlearn}
\bibfield{author}{\bibinfo{person}{Guillaume Lema{{\^i}}tre},
  \bibinfo{person}{Fernando Nogueira}, {and} \bibinfo{person}{Christos~K.
  Aridas}.} \bibinfo{year}{2017}\natexlab{}.
\newblock \showarticletitle{Imbalanced-learn: A Python Toolbox to Tackle the
  Curse of Imbalanced Datasets in Machine Learning}.
\newblock \bibinfo{journal}{\emph{Journal of Machine Learning Research}}
  \bibinfo{volume}{18}, \bibinfo{number}{17} (\bibinfo{year}{2017}),
  \bibinfo{pages}{1--5}.
\newblock
\urldef\tempurl%
\url{http://jmlr.org/papers/v18/16-365.html}
\showURL{%
\tempurl}


\bibitem[Lin et~al\mbox{.}(2014)]%
        {COCO}
\bibfield{author}{\bibinfo{person}{Tsung{-}Yi Lin}, \bibinfo{person}{Michael
  Maire}, \bibinfo{person}{Serge~J. Belongie}, {and} \bibinfo{person}{Others}.}
  \bibinfo{year}{2014}\natexlab{}.
\newblock \showarticletitle{Microsoft {COCO:} Common Objects in Context}.
\newblock \bibinfo{journal}{\emph{CoRR}}  \bibinfo{volume}{abs/1405.0312}
  (\bibinfo{year}{2014}).
\newblock
\showeprint[arXiv]{1405.0312}


\bibitem[Maposa et~al\mbox{.}(2010)]%
        {MANOVA}
\bibfield{author}{\bibinfo{person}{Daniel Maposa}, \bibinfo{person}{Edinah
  Mudimu}, {and} \bibinfo{person}{Olina Ngwenya}.}
  \bibinfo{year}{2010}\natexlab{}.
\newblock \showarticletitle{A multivariate analysis of variance (MANOVA) of the
  performance of sorghum lines in different agro-ecological regions of
  Zimbabwe}.
\newblock \bibinfo{journal}{\emph{African Journal of Agricultural Research}}
  \bibinfo{volume}{5} (\bibinfo{date}{02} \bibinfo{year}{2010}),
  \bibinfo{pages}{196--203}.
\newblock


\bibitem[Martinez-Pecino and Garcia-Gavil{\'a}n(2019)]%
        {martinez2019likes}
\bibfield{author}{\bibinfo{person}{Roberto Martinez-Pecino} {and}
  \bibinfo{person}{Marta Garcia-Gavil{\'a}n}.} \bibinfo{year}{2019}\natexlab{}.
\newblock \showarticletitle{Likes and problematic Instagram use: the moderating
  role of self-esteem}.
\newblock \bibinfo{journal}{\emph{Cyberpsychology, Behavior, and Social
  Networking}} \bibinfo{volume}{22}, \bibinfo{number}{6}
  (\bibinfo{year}{2019}), \bibinfo{pages}{412--416}.
\newblock


\bibitem[Maslowska et~al\mbox{.}(2019)]%
        {maslowska2019online}
\bibfield{author}{\bibinfo{person}{Ewa Maslowska}, \bibinfo{person}{Su~Jung
  Kim}, \bibinfo{person}{Edward~C Malthouse}, {and} \bibinfo{person}{Vijay
  Viswanathan}.} \bibinfo{year}{2019}\natexlab{}.
\newblock \showarticletitle{Online reviews as customers’ dialogues with and
  about brands}.
\newblock In \bibinfo{booktitle}{\emph{Handbook of research on customer
  engagement}}. \bibinfo{publisher}{Edward Elgar Publishing}.
\newblock


\bibitem[Mazloom et~al\mbox{.}(2018)]%
        {mazloom2018category}
\bibfield{author}{\bibinfo{person}{Masoud Mazloom}, \bibinfo{person}{Iliana
  Pappi}, {and} \bibinfo{person}{Marcel Worring}.}
  \bibinfo{year}{2018}\natexlab{}.
\newblock \showarticletitle{Category specific post popularity prediction}. In
  \bibinfo{booktitle}{\emph{International Conference on Multimedia Modeling}}.
  Springer, \bibinfo{pages}{594--607}.
\newblock


\bibitem[Mazloom et~al\mbox{.}(2016)]%
        {mazloom2016multimodal}
\bibfield{author}{\bibinfo{person}{Masoud Mazloom}, \bibinfo{person}{Robert
  Rietveld}, \bibinfo{person}{Stevan Rudinac}, \bibinfo{person}{Marcel
  Worring}, {and} \bibinfo{person}{Willemijn Van~Dolen}.}
  \bibinfo{year}{2016}\natexlab{}.
\newblock \showarticletitle{Multimodal popularity prediction of brand-related
  social media posts}. In \bibinfo{booktitle}{\emph{Proceedings of the 24th ACM
  international conference on Multimedia}}. \bibinfo{pages}{197--201}.
\newblock


\bibitem[Mehrabian and Russell(1974)]%
        {mehrabian1974approach}
\bibfield{author}{\bibinfo{person}{Albert Mehrabian} {and}
  \bibinfo{person}{James~A Russell}.} \bibinfo{year}{1974}\natexlab{}.
\newblock \bibinfo{booktitle}{\emph{An approach to environmental psychology.}}
\newblock \bibinfo{publisher}{the MIT Press}.
\newblock


\bibitem[Molnar(2020)]%
        {molnar2020interpretable}
\bibfield{author}{\bibinfo{person}{Christoph Molnar}.}
  \bibinfo{year}{2020}\natexlab{}.
\newblock \bibinfo{booktitle}{\emph{Interpretable machine learning}}.
\newblock \bibinfo{publisher}{Lulu. com}.
\newblock


\bibitem[mowshon(2022)]%
        {agegender}
\bibfield{author}{\bibinfo{person}{mowshon}.} \bibinfo{year}{2022}\natexlab{}.
\newblock \bibinfo{title}{Age and Gender}.
\newblock
\newblock
\urldef\tempurl%
\url{https://github.com/mowshon/age-and-gender}
\showURL{%
\tempurl}


\bibitem[Myers et~al\mbox{.}(2010)]%
        {statisticbook}
\bibfield{author}{\bibinfo{person}{J.L. Myers}, \bibinfo{person}{A. Well},
  {and} \bibinfo{person}{R.F. Lorch}.} \bibinfo{year}{2010}\natexlab{}.
\newblock \bibinfo{booktitle}{\emph{Research Design and Statistical Analysis}}.
\newblock \bibinfo{publisher}{Routledge}.
\newblock
\showISBNx{9780805864311}
\showLCCN{2009032606}


\bibitem[Newberry(2022)]%
        {instagram_growth}
\bibfield{author}{\bibinfo{person}{Christina Newberry}.}
  \bibinfo{year}{2022}\natexlab{}.
\newblock \bibinfo{title}{12 Foolproof Instagram Growth Strategies for 2022}.
\newblock
\newblock
\urldef\tempurl%
\url{https://blog.hootsuite.com/instagram-growth/}
\showURL{%
\tempurl}
\newblock
\shownote{Accessed: August 2022}.


\bibitem[Niu et~al\mbox{.}(2021)]%
        {niu2021spice}
\bibfield{author}{\bibinfo{person}{Chuang Niu}, \bibinfo{person}{Hongming
  Shan}, {and} \bibinfo{person}{Ge Wang}.} \bibinfo{year}{2021}\natexlab{}.
\newblock \showarticletitle{Spice: Semantic pseudo-labeling for image
  clustering}.
\newblock \bibinfo{journal}{\emph{arXiv preprint arXiv:2103.09382}}
  (\bibinfo{year}{2021}).
\newblock


\bibitem[notAI.tech(2022)]%
        {nudenet}
\bibfield{author}{\bibinfo{person}{notAI.tech}.}
  \bibinfo{year}{2022}\natexlab{}.
\newblock \bibinfo{title}{NudeNet: Neural Nets for Nudity Classification,
  Detection and Selective Censoring}.
\newblock
\newblock
\urldef\tempurl%
\url{https://github.com/notAI-tech/NudeNet}
\showURL{%
\tempurl}
\newblock
\shownote{Accessed: August 2022}.


\bibitem[Paszke et~al\mbox{.}(2019)]%
        {NEURIPS20199015}
\bibfield{author}{\bibinfo{person}{Adam Paszke}, \bibinfo{person}{Sam Gross},
  \bibinfo{person}{Massa}, {et~al\mbox{.}}} \bibinfo{year}{2019}\natexlab{}.
\newblock \showarticletitle{PyTorch: An Imperative Style, High-Performance Deep
  Learning Library}.
\newblock In \bibinfo{booktitle}{\emph{Advances in Neural Information
  Processing Systems 32}}, \bibfield{editor}{\bibinfo{person}{H.~Wallach},
  \bibinfo{person}{H.~Larochelle}, \bibinfo{person}{A.~Beygelzimer},
  \bibinfo{person}{F.~d\textquotesingle Alch\'{e}-Buc},
  \bibinfo{person}{E.~Fox}, {and} \bibinfo{person}{R.~Garnett}} (Eds.).
  \bibinfo{publisher}{Curran Associates, Inc.}, \bibinfo{pages}{8024--8035}.
\newblock


\bibitem[Pedregosa et~al\mbox{.}(2011)]%
        {sklearn}
\bibfield{author}{\bibinfo{person}{F. Pedregosa}, \bibinfo{person}{G.
  Varoquaux}, {et~al\mbox{.}}} \bibinfo{year}{2011}\natexlab{}.
\newblock \showarticletitle{Scikit-learn: Machine Learning in {P}ython}.
\newblock \bibinfo{journal}{\emph{Journal of Machine Learning Research}}
  \bibinfo{volume}{12} (\bibinfo{year}{2011}), \bibinfo{pages}{2825--2830}.
\newblock


\bibitem[Purba et~al\mbox{.}(2021)]%
        {purba2021instagram}
\bibfield{author}{\bibinfo{person}{Kristo~Radion Purba}, \bibinfo{person}{David
  Asirvatham}, {and} \bibinfo{person}{Raja~Kumar Murugesan}.}
  \bibinfo{year}{2021}\natexlab{}.
\newblock \showarticletitle{Instagram post popularity trend analysis and
  prediction using hashtag, image assessment, and user history features.}
\newblock \bibinfo{journal}{\emph{Int. Arab J. Inf. Technol.}}
  \bibinfo{volume}{18}, \bibinfo{number}{1} (\bibinfo{year}{2021}),
  \bibinfo{pages}{85--94}.
\newblock


\bibitem[Quinlan(1986)]%
        {quinlan1986induction}
\bibfield{author}{\bibinfo{person}{J.~Ross Quinlan}.}
  \bibinfo{year}{1986}\natexlab{}.
\newblock \showarticletitle{Induction of decision trees}.
\newblock \bibinfo{journal}{\emph{Machine learning}} \bibinfo{volume}{1},
  \bibinfo{number}{1} (\bibinfo{year}{1986}), \bibinfo{pages}{81--106}.
\newblock


\bibitem[Ransaw(2021)]%
        {whyshare}
\bibfield{author}{\bibinfo{person}{Rosalyn Ransaw}.}
  \bibinfo{year}{2021}\natexlab{}.
\newblock \bibinfo{title}{The Psychology Behind Why We Share on Social Media}.
\newblock
\newblock
\urldef\tempurl%
\url{https://www.shutterstock.com/blog/the-psychology-behind-why-we-share-on-social-media}
\showURL{%
\tempurl}
\newblock
\shownote{Accessed: August 2022}.


\bibitem[Reimers and Gurevych(2019)]%
        {reimers2019sentence}
\bibfield{author}{\bibinfo{person}{Nils Reimers} {and} \bibinfo{person}{Iryna
  Gurevych}.} \bibinfo{year}{2019}\natexlab{}.
\newblock \showarticletitle{Sentence-BERT: Sentence Embeddings using Siamese
  BERT-Networks}.
\newblock  (\bibinfo{year}{2019}), \bibinfo{pages}{3982--3992}.
\newblock


\bibitem[Riis et~al\mbox{.}(2020)]%
        {riis2020limits}
\bibfield{author}{\bibinfo{person}{Christoffer Riis},
  \bibinfo{person}{Damian~Konrad Kowalczyk}, {and} \bibinfo{person}{Lars~Kai
  Hansen}.} \bibinfo{year}{2020}\natexlab{}.
\newblock \showarticletitle{On the limits to multi-modal popularity prediction
  on instagram--a new robust, efficient and explainable baseline}.
\newblock \bibinfo{journal}{\emph{arXiv preprint arXiv:2004.12482}}
  (\bibinfo{year}{2020}).
\newblock


\bibitem[Rudin(2019)]%
        {rudin2019stop}
\bibfield{author}{\bibinfo{person}{Cynthia Rudin}.}
  \bibinfo{year}{2019}\natexlab{}.
\newblock \showarticletitle{Stop explaining black box machine learning models
  for high stakes decisions and use interpretable models instead}.
\newblock \bibinfo{journal}{\emph{Nature Machine Intelligence}}
  \bibinfo{volume}{1}, \bibinfo{number}{5} (\bibinfo{year}{2019}),
  \bibinfo{pages}{206--215}.
\newblock


\bibitem[Sharma et~al\mbox{.}(2022)]%
        {sharma2022detecting}
\bibfield{author}{\bibinfo{person}{Shivam Sharma}, \bibinfo{person}{Firoj
  Alam}, \bibinfo{person}{Md Akhtar}, \bibinfo{person}{Dimitar Dimitrov},
  \bibinfo{person}{Giovanni Da~San Martino}, \bibinfo{person}{Hamed Firooz},
  \bibinfo{person}{Alon Halevy}, \bibinfo{person}{Fabrizio Silvestri},
  \bibinfo{person}{Preslav Nakov}, \bibinfo{person}{Tanmoy Chakraborty},
  {et~al\mbox{.}}} \bibinfo{year}{2022}\natexlab{}.
\newblock \showarticletitle{Detecting and Understanding Harmful Memes: A
  Survey}.
\newblock \bibinfo{journal}{\emph{arXiv preprint arXiv:2205.04274}}
  (\bibinfo{year}{2022}).
\newblock


\bibitem[Sherman and Quester(2006)]%
        {sherman2006influence}
\bibfield{author}{\bibinfo{person}{Claire Sherman} {and}
  \bibinfo{person}{Pascale Quester}.} \bibinfo{year}{2006}\natexlab{}.
\newblock \showarticletitle{The influence of product/nudity congruence on
  advertising effectiveness}.
\newblock \bibinfo{journal}{\emph{Journal of Promotion Management}}
  \bibinfo{volume}{11}, \bibinfo{number}{2-3} (\bibinfo{year}{2006}),
  \bibinfo{pages}{61--89}.
\newblock


\bibitem[Simpson(1949)]%
        {simpson1949measurement}
\bibfield{author}{\bibinfo{person}{Edward~H Simpson}.}
  \bibinfo{year}{1949}\natexlab{}.
\newblock \showarticletitle{Measurement of diversity}.
\newblock \bibinfo{journal}{\emph{nature}} \bibinfo{volume}{163},
  \bibinfo{number}{4148} (\bibinfo{year}{1949}), \bibinfo{pages}{688--688}.
\newblock


\bibitem[Statusbrew(2021)]%
        {instaalg2022}
\bibfield{author}{\bibinfo{person}{Statusbrew}.}
  \bibinfo{year}{2021}\natexlab{}.
\newblock \bibinfo{title}{Instagram Algorithm 2022: How To Conquer It}.
\newblock
\newblock
\urldef\tempurl%
\url{https://statusbrew.com/insights/instagram-algorithm/}
\showURL{%
\tempurl}
\newblock
\shownote{Accessed: August 2022}.


\bibitem[Tabachnick et~al\mbox{.}(2007)]%
        {tabachnick2007using}
\bibfield{author}{\bibinfo{person}{Barbara~G Tabachnick},
  \bibinfo{person}{Linda~S Fidell}, {and} \bibinfo{person}{Jodie~B Ullman}.}
  \bibinfo{year}{2007}\natexlab{}.
\newblock \bibinfo{booktitle}{\emph{Using multivariate statistics}}.
  Vol.~\bibinfo{volume}{5}.
\newblock \bibinfo{publisher}{pearson Boston, MA}.
\newblock


\bibitem[Tricomi et~al\mbox{.}(2022)]%
        {tricomi2022we}
\bibfield{author}{\bibinfo{person}{Pier~Paolo Tricomi}, \bibinfo{person}{Sousan
  Tarahomi}, \bibinfo{person}{Christian Cattai}, \bibinfo{person}{Francesco
  Martini}, {and} \bibinfo{person}{Mauro Conti}.}
  \bibinfo{year}{2022}\natexlab{}.
\newblock \showarticletitle{Are We All in a Truman Show? Spotting Instagram
  Crowdturfing through Self-Training}.
\newblock \bibinfo{journal}{\emph{arXiv preprint arXiv:2206.12904}}
  (\bibinfo{year}{2022}).
\newblock


\bibitem[Valdez and Mehrabian(1994)]%
        {valdez1994effects}
\bibfield{author}{\bibinfo{person}{Patricia Valdez} {and}
  \bibinfo{person}{Albert Mehrabian}.} \bibinfo{year}{1994}\natexlab{}.
\newblock \showarticletitle{Effects of color on emotions.}
\newblock \bibinfo{journal}{\emph{Journal of experimental psychology: General}}
  \bibinfo{volume}{123}, \bibinfo{number}{4} (\bibinfo{year}{1994}),
  \bibinfo{pages}{394}.
\newblock


\bibitem[Zarei et~al\mbox{.}(2020)]%
        {zarei2020impersonators}
\bibfield{author}{\bibinfo{person}{Koosha Zarei}, \bibinfo{person}{Reza
  Farahbakhsh}, {and} \bibinfo{person}{No{\"e}l Crespi}.}
  \bibinfo{year}{2020}\natexlab{}.
\newblock \showarticletitle{How impersonators exploit Instagram to generate
  fake engagement?}. In \bibinfo{booktitle}{\emph{ICC 2020-2020 IEEE
  International Conference on Communications (ICC)}}. IEEE,
  \bibinfo{pages}{1--6}.
\newblock


\bibitem[Zarrinkalam et~al\mbox{.}(2020)]%
        {zarrinkalam2020extracting}
\bibfield{author}{\bibinfo{person}{Fattane Zarrinkalam},
  \bibinfo{person}{Stefano Faralli}, \bibinfo{person}{Guangyuan Piao},
  \bibinfo{person}{Ebrahim Bagheri}, {et~al\mbox{.}}}
  \bibinfo{year}{2020}\natexlab{}.
\newblock \showarticletitle{Extracting, mining and predicting users’
  interests from social media}.
\newblock \bibinfo{journal}{\emph{Foundations and Trends{\textregistered} in
  Information Retrieval}} \bibinfo{volume}{14}, \bibinfo{number}{5}
  (\bibinfo{year}{2020}), \bibinfo{pages}{445--617}.
\newblock


\bibitem[Zeng et~al\mbox{.}(2016)]%
        {zeng2016effective}
\bibfield{author}{\bibinfo{person}{Min Zeng}, \bibinfo{person}{Beiji Zou},
  \bibinfo{person}{Faran Wei}, \bibinfo{person}{Xiyao Liu}, {and}
  \bibinfo{person}{Lei Wang}.} \bibinfo{year}{2016}\natexlab{}.
\newblock \showarticletitle{Effective prediction of three common diseases by
  combining SMOTE with Tomek links technique for imbalanced medical data}. In
  \bibinfo{booktitle}{\emph{2016 IEEE International Conference of Online
  Analysis and Computing Science (ICOACS)}}. IEEE, \bibinfo{pages}{225--228}.
\newblock


\bibitem[Zhang et~al\mbox{.}(2018b)]%
        {zhang2018trending}
\bibfield{author}{\bibinfo{person}{Lei Zhang}, \bibinfo{person}{Paul Jones},
  \bibinfo{person}{Kent~Aaron Otis}, \bibinfo{person}{Jonathan Gale}, {and}
  \bibinfo{person}{Evelyn Chan}.} \bibinfo{year}{2018}\natexlab{b}.
\newblock \bibinfo{title}{Trending topic extraction from social media}.
\newblock
\newblock
\newblock
\shownote{US Patent 10,095,686}.


\bibitem[Zhang et~al\mbox{.}(2018a)]%
        {zhang2018become}
\bibfield{author}{\bibinfo{person}{Zhongping Zhang}, \bibinfo{person}{Tianlang
  Chen}, \bibinfo{person}{Zheng Zhou}, \bibinfo{person}{Jiaxin Li}, {and}
  \bibinfo{person}{Jiebo Luo}.} \bibinfo{year}{2018}\natexlab{a}.
\newblock \showarticletitle{How to become instagram famous: Post popularity
  prediction with dual-attention}. In \bibinfo{booktitle}{\emph{2018 IEEE
  international conference on big data (big data)}}. IEEE,
  \bibinfo{pages}{2383--2392}.
\newblock


\bibitem[Zhou et~al\mbox{.}(2017)]%
        {places365}
\bibfield{author}{\bibinfo{person}{Bolei Zhou}, \bibinfo{person}{Agata
  Lapedriza}, \bibinfo{person}{Aditya Khosla}, \bibinfo{person}{Aude Oliva},
  {and} \bibinfo{person}{Antonio Torralba}.} \bibinfo{year}{2017}\natexlab{}.
\newblock \showarticletitle{Places: A 10 million Image Database for Scene
  Recognition}.
\newblock \bibinfo{journal}{\emph{IEEE Transactions on Pattern Analysis and
  Machine Intelligence}} (\bibinfo{year}{2017}).
\newblock


\bibitem[Zohourian et~al\mbox{.}(2018)]%
        {zohourian2018popularity}
\bibfield{author}{\bibinfo{person}{Alireza Zohourian}, \bibinfo{person}{Hedieh
  Sajedi}, {and} \bibinfo{person}{Arefeh Yavary}.}
  \bibinfo{year}{2018}\natexlab{}.
\newblock \showarticletitle{Popularity prediction of images and videos on
  Instagram}. In \bibinfo{booktitle}{\emph{2018 4th International Conference on
  Web Research (ICWR)}}. IEEE, \bibinfo{pages}{111--117}.
\newblock


\bibitem[Zwillinger and Kokoska(1999)]%
        {spearman}
\bibfield{author}{\bibinfo{person}{Daniel Zwillinger} {and}
  \bibinfo{person}{Stephen Kokoska}.} \bibinfo{year}{1999}\natexlab{}.
\newblock \bibinfo{booktitle}{\emph{CRC standard probability and statistics
  tables and formulae}}.
\newblock \bibinfo{publisher}{Crc Press}.
\newblock


\end{thebibliography}
